\documentclass{aa}
\usepackage{soul}
\usepackage{color}

\usepackage{graphicx}
\usepackage{txfonts}
\usepackage{lipsum}
\usepackage{subcaption}         % necessary for continued figures, example in section 3
\usepackage{lscape}             % to rotate a single page table, example in appendix.
\usepackage{placeins}           % useful with \FloatBarrier, to keep 
\usepackage{hyperref}
\begin{document}

%%%%%%%%%%%%%%%%%%%%%%%%%%%%%%%%%%%%%%%%
% if you use custom commands in your title,
% ensure to check your title when submitting!
%%%%%%%%%%%%%%%%%%%%%%%%%%%%%%%%%%%%%%%%

   \title{Tracking the evolution of sub-Neptunes.}

   \subtitle{I. Gyrochronology of planet hosting stars.}
   
%%%%%%%%%%%%%%%%%%%%%%%%%%%%%%%%%%%%%%%%
% Please separate each author with the \and command
%
% Use the \corrauth to provide the corresponding
% author address. It will be automatically inserted as 
% footnote in the PDF output.
%
% Please DO NOT include ORCIDs next to author names.
% Instead, please provide an active address for each coauthor:
% it will be automatically extracted by EDPS editorial system, 
% and co-authors will be be able to authenticate their ORCID.
%
% Only authenticated ORCIDs will be taken into account.
% ORCIDs included here will be removed.
%%%%%%%%%%%%%%%%%%%%%%%%%%%%%%%%%%%%%%%%

   \author{A.C.~Lanzafame\inst{\ref{inst:0001},\ref{inst:0002}}\corrauth{a.lanzafame@unict.it}
        \and
        V.~Drago\inst{\ref{inst:0001}}\email{vincedrago89@gmail.com}
        \and
        N.~Greco\inst{\ref{inst:0001},\ref{inst:0002}}\email{nino.greco@inaf.it}
        \and
        I.~Pagano\inst{\ref{inst:0003}}\email{isabella.pagano@inaf.it}
        \and
        G.~Scandariato\inst{\ref{inst:0002}}\email{gaetano.scandariato@inaf.it}
        \and
        F.~Spada\inst{\ref{inst:0001}}\email{federico.spada@dfa.unict.it}
        }
   \institute{
     Dipartimento di Fisica e Astronomia ``Ettore Majorana'', Universit\`{a} di Catania, Via S. Sofia 64, 95123 Catania, Italy\relax\label{inst:0001}
   \and 
     INAF - Osservatorio Astrofisico di Catania, Via S. Sofia 78, 95123 Catania, Italy\relax \label{inst:0002}
   \and
    INAF, Istituto Nazionale di Astrofisica, Viale del Parco Mellini 84, 00136 Roma, Italy\relax \label{inst:0003}
   }
      
   %\date{Received September 30, 20XX}
   \date{Received \today}

\abstract{
Stellar age is fundamental for reconstructing the evolution of planetary systems, but remains difficult to determine for unevolved main-sequence stars. Stellar rotation provides a sensitive age diagnostic in this regime and can therefore establish the temporal framework required to investigate the evolution of close-in planets.
}{
This work is the first in a series aimed at tracing the evolution of the sub-Neptune population through a homogeneous determination of the ages of their host stars.
}{
We compiled rotation periods and stellar parameters for 4639 confirmed exoplanet hosts. Gyrochronological ages were inferred using the two-zone rotational evolution model of \citet{SL:26}. We adopted a Bayesian approach that simultaneously compares the observed rotation period and de-reddened Gaia $(G_{\rm BP}-G_{\rm RP})$ colour with model predictions, using literature stellar masses as Gaussian priors. The rotational tracks were extended to the terminal-age main sequence to characterise stars beyond the empirically calibrated age domain.
}{
We obtain gyrochronological constraints for 2204 exoplanet-host stars. For stars with well-constrained posterior distributions, typical formal age uncertainties are $10$--$15\%$, providing substantially greater age resolution than is generally attainable from isochrone fitting for unevolved main-sequence stars. The inferred stellar masses are consistent with the literature values used as priors. The model effective temperature agrees well with the determinations in the independent literature, providing an additional consistency check on the inferred stellar properties. Ages outside the open-cluster calibration range should be regarded as model-dependent estimates.
}{ 
The combination of sample size, homogeneity, and age precision provides a powerful data set for studying the temporal evolution of exoplanet populations around main-sequence stars. It establishes the age framework for reconstructing the evolution of the sub-Neptune population and the emergence of the radius valley. The stellar parameters, their provenance, and the gyrochronological results are publicly available through the CDS.
}


   \maketitle
%%%%%%%%%%%%%%%%%%%%%%%%%%%%%%%%%%%%%%%%%%%%%%%%%%%%%%%%%%%%%%

\nolinenumbers

\section{Introduction}
\label{sec:introduction}

Time is a fundamental variable in the evolution of planetary systems. The architecture and physical properties of a planetary system observed at a given epoch result from processes operating over a wide range of timescales, from the dispersal of the protoplanetary disc during the first few Myr to the secular evolution of planets and their host stars over several Gyr. Establishing how planetary populations evolve therefore requires reliable stellar ages. This remains one of the main limitations in observational studies of exoplanet evolution, particularly for the low-mass main-sequence stars that host the majority of known planetary systems.

The problem is especially relevant for close-in sub-Neptunes. Planets with sizes between those of the Earth and Neptune constitute one of the most common outcomes of planet formation, despite having no direct analogue in the Solar System. Their radius distribution exhibits a pronounced deficit separating predominantly rocky super-Earths from larger sub-Neptunes, usually referred to as the radius valley \citep{2017AJ....154..109F,2018AJ....156..264F}. Its existence is generally interpreted as evidence that the present-day radii of close-in planets are strongly affected by post-formation evolution. Atmospheric mass loss driven by stellar high-energy irradiation \citep[e.g.][]{2007ApJ...669..585L,2013ApJ...775..105O} and cooling-powered mass loss from the planetary core and envelope \citep[e.g.][]{2018MNRAS.479.4391G,2020MNRAS.491.2885G} provide two leading explanations for the observed separation between the two populations. Other processes associated with planet formation and the properties of the primordial atmosphere may also contribute to shaping the observed distribution.

Discriminating among these mechanisms requires more than characterising the radius distribution at the present epoch. Their characteristic timescales and predicted evolutionary paths differ, and the relative populations of super-Earths and sub-Neptunes are therefore expected to evolve with time. The age dependence of the radius valley, and more generally of the joint distribution of planetary radius, orbital properties, and incident stellar flux, contains information that cannot be recovered from a time-integrated population alone. Observationally reconstructing this evolution requires a sufficiently large sample of planets whose host stars are placed on a common and precise age scale.

Obtaining such an age scale is challenging. Stellar age is not directly observable and must instead be inferred from quantities that evolve with time. Isochrone fitting \cite{1962ApJ...135..349S,1964ApJ...140..544D}, is a powerful technique for evolved stars, whose positions in the Hertzsprung--Russell diagram change appreciably with age, but provides relatively weak constraints for unevolved low-mass main-sequence stars because their luminosities and effective temperatures evolve slowly. Asteroseismology can provide accurate ages but is applicable only to a comparatively restricted sample of stars with suitable observations. Chemical clocks and abundance-based methods can be powerful for selected stellar populations, but their calibration and applicability depend on stellar properties and Galactic chemical evolution. Consequently, ages collected from the literature for exoplanet hosts are often derived using different techniques, assumptions, and stellar models, introducing substantial heterogeneity when they are combined for population studies.

Rotation offers a particularly powerful age diagnostic for cool main-sequence stars. Stars with convective envelopes lose angular momentum through magnetised stellar winds, producing a systematic evolution of surface rotation with time. The empirical connection between stellar rotation and age was established from observations of the Sun and open clusters \citep{1972ApJ...171..565S} and forms the basis of gyrochronology \citep{2003ApJ...586..464B,2007ApJ...669.1167B}. Open-cluster observations reveal that, from an initially broad distribution of rotation rates, stars progressively converge onto a well-defined slow-rotator sequence whose subsequent evolution depends primarily on stellar mass and age. The emergence of this sequence is already apparent, particularly among the higher-mass cool stars, in the $\simeq0.1$\,Gyr-old Pleiades cluster, while progressively lower-mass stars converge at later ages. This mass-dependent convergence makes rotation particularly useful as a clock for main-sequence stars, precisely where conventional isochrone ages become poorly constrained.

Modern rotational evolution models go beyond simple separable period--colour--age relations by describing the physical evolution of stellar angular momentum. In particular, two-zone models account for the exchange of angular momentum between the radiative core and convective envelope, together with angular-momentum loss through magnetised winds. Such models reproduce the mass-dependent evolution of the slow-rotator sequences observed in open clusters and provide a physically motivated framework for inferring stellar ages \citep[e.g.][]{2015A&A...584A..30L}. Their empirical calibration is nevertheless limited by the age range sampled by clusters with reliable rotational measurements. At older ages, observations of field stars indicate departures from standard spin-down prescriptions, including the possible onset of weakened magnetic braking at sufficiently large Rossby number \citep{2016Natur.529..181V}. These limitations need to be explicitly taken into account when gyrochronology is applied beyond the domain directly constrained by cluster observations.

For exoplanet studies, gyrochronology has two particularly attractive properties. First, the rotation period can often provide much greater age discrimination than the position of an unevolved star in the Hertzsprung--Russell diagram. Second, applying a single rotational evolution model to the entire sample provides a homogeneous relative age scale, avoiding much of the methodological heterogeneity inherent in compilations of literature ages. 
Rotation periods are now available for a substantial number of planet-hosting stars, both from large space-based and ground-based photometric surveys and from targeted studies of individual planetary systems. The latter are particularly valuable because dedicated analyses of planet hosts can provide rotation measurements for stars not included in homogeneous survey catalogues. Together with the precise and homogeneous astrometry and photometry provided by \textit{Gaia}, these measurements make it possible to apply gyrochronology to a large and heterogeneous population of planet-hosting stars.

This paper is the first in the \textit{Tracking the evolution of sub-Neptunes} series, whose aim is to reconstruct observationally how the properties and population of close-in sub-Neptunes evolve with time. The purpose of the present work is to establish the stellar-age foundation for this analysis. We compile stellar parameters and reliable rotation-period measurements for confirmed exoplanet hosts and derive their ages using the two-zone rotational evolution model of \citet{SL:26}. The model is calibrated against open-cluster rotation sequences over the mass range $0.4$--$1.25\,\mathcal{M}_{\odot}$ and is extended here to provide a continuous age scale up to the terminal-age main sequence, while explicitly identifying estimates that lie outside the empirically calibrated domain.

Rather than deriving ages from rotation alone, we formulate the inference in a Bayesian framework in which the observed rotation period and de-reddened Gaia $(G_{\rm BP}-G_{\rm RP})$ colour are compared simultaneously with the rotational and stellar evolutionary models, while independent stellar masses from the literature are included as prior information. This approach allows observational uncertainties, the intrinsic width of the rotational sequence, and uncertainties in stellar mass to be propagated into the age posterior. It also provides model predictions for stellar properties that can be compared with independent literature determinations as a consistency check.

The resulting catalogue provides gyrochronological constraints for 2204 confirmed exoplanet-host stars drawn from an initial compilation of 4716 systems. For stars within the domain in which rotational evolution is well constrained, the method provides substantially greater age precision than is generally achievable from isochrone fitting for unevolved main-sequence stars, while maintaining a common age scale across the sample. The catalogue is designed not merely as a compilation of stellar ages, but as the temporal framework for the subsequent papers in this series. These ages will be used to investigate the evolution of the sub-Neptune population and, in particular, to trace how the radius distribution and radius valley develop with stellar age, thereby providing direct observational constraints on the physical mechanisms responsible for shaping close-in planetary systems.

The paper is organised as follows. Section~\ref{sec:data} describes the compilation and selection of the stellar sample and rotation-period measurements. Section~\ref{sec:model} presents the rotational evolution model and its extension, while Sect.~\ref{sec:inference} describes the Bayesian inference of stellar ages. We then present the resulting gyrochronological catalogue and assess the consistency of the inferred stellar parameters with independent literature determinations in Sect.~\ref{sec:results}. Finally, we summarise our results and discuss the applicability and limitations of the resulting age scale in Sect.~\ref{sec:conclusions}.

%%%%%%%%%%%%%%%%%%%%%%%%%%%%%%%%%%%%%%%%%%%%%%%%%%%%%%%%%%%%%%

\section{Data}
\label{sec:data}

The gyrochronological analysis described in Sect.\,\ref{sec:model} requires, for each
star, three observables: a surface rotation period $P_\star$, the de-reddened colour index
$C_\star \equiv (G_{\rm BP}-G_{\rm RP})$, and an independent estimate of the stellar mass
$\mathcal{M}_0$ used as a prior (Sect.\,\ref{sec:inference}); we additionally compile the
effective temperature and metallicity, which do not enter the likelihood. These quantities
are drawn from a heterogeneous compilation of literature and survey-archive measurements and
reduced, star by star, to a single set of recommended values by the
selection procedure described below. 

\subsection{The host-star sample}
\label{sec:catalogue}

For each confirmed planet-hosting star we collected, from the public literature and survey
archives, all the available measurements of the stellar quantities required by the analysis:
the surface rotation period, the {\it Gaia} colour, the mass, and, as auxiliary parameters,
the effective temperature and metallicity. The list of confirmed planets and their host
stars is taken from the Planetary Systems table of the NASA Exoplanet Archive
\citep[NEA;][]{Akeson2013}, retrieved on 2026 June~19\footnote{\url{https://exoplanetarchive.ipac.caltech.edu/}};
the astrometric and
photometric data, in particular the colour index, are drawn from the third {\it Gaia} data
release \citep[Gaia~DR3;][]{GaiaDR3}; and the rotation periods are assembled from dedicated
catalogues and individual studies. The
reduction of these multiple values to a single recommended estimate per quantity is summarised
in the following subsections and described in full, branch by branch, in Appendix~\ref{app:recon}; the complete per-source compilation of exoplanet
and host-star data is presented in a dedicated companion paper \citep[][hereafter Paper II]{PaperII}.

Of the $4716$ confirmed planet-hosting stars in the NEA compilation,
the subsample relevant to the present work (those with at least one published
surface-rotation measurement) consists of $2874$ stars, with a total of $\sim$$10\,700$
rotation-period values drawn from about two dozen dedicated catalogues in addition to
numerous individual system studies.
These
span the {\it Kepler}, K2, and TESS photometric surveys, ground-based photometric monitoring,
and spectroscopic chromospheric-activity time series, and were obtained with a variety of
period-finding techniques (auto-correlation, Lomb--Scargle and generalised Lomb--Scargle
periodograms, Gaussian-process regression, wavelet and gradient-of-power-spectrum methods,
and deep learning); the most heavily used sources include
\citet{McQuillan2013,Mazeh2015,David2021,Santos2019,Santos2021,WalkowiczBasri2013,Angus2018,Reinhold2013,Reinhold2023,Kamai2025,Boyle2026}.

\subsection{Rotation periods}
\label{sec:rotation}

For a large fraction of the sample several independent rotation periods are available, and
they are sometimes mutually inconsistent. The disagreements are only occasionally genuine:
most often they form \emph{harmonic families} ($P$, $P/2$, $2P$, $P/3$, and higher harmonics) produced by different
period-finding techniques operating on the same light curve, with a contamination of
\emph{non-rotational} periodicities (e.g. asteroseismic oscillations of evolved stars or $\delta$\,Scuti
pulsations of early-type stars) reported in the literature as rotation periods. We therefore reduce,
for every star, the discordant literature values to a \emph{single recommended period}, chosen from the available
measurements.

Before any selection, a few cleaning steps are applied: measurements derived from the
projected rotational velocity $v\sin i$, predicted from an activity--rotation relation, or
flagged as upper limits are removed; a value re-tabulated unchanged by several catalogues is counted once;
and, where a compilation catalogue mixes its own measurements with re-quotes of earlier work,
only the periods it actually endorses are kept (Appendix~\ref{app:recon}).

Two physical limits then take precedence over any statistical argument. 
First, a measurement
shorter than the critical (break-up) period
is rotationally impossible and is discarded; a star 
whose measured periods all fall below that limit is classified as having a non-rotational signal and is excluded.
Second, the projected rotational velocity sets a hard upper limit on the period
($P_{\rm max}\propto (R/R_\odot)/v\sin i$, since $\sin i \le 1$): within the selection, a value exceeding
$1.3\,P_{\rm max}$ is rejected in favour of a shorter, $v\sin i$-consistent one where available, only above a $4\,{\rm km\,s^{-1}}$ floor, below which
unmodelled macroturbulence inflates the measured $v\sin i$ and makes the ceiling unreliable.
Both relations and the calibration floor are detailed in Appendix~\ref{app:guards:vetoes}.
Of the $2874$ stars, $158$ lose all their measurements to these cleaning steps; the remaining $2716$ retain at least one usable period.

The surviving measurements of each star are then reduced to a single recommended period
by a documented, deterministic procedure: discordant sets are resolved by an ordered sequence
of physical and statistical discriminants that identify which measurements trace the true
rotation, described in full in Appendix~\ref{app:levers}. 

The recommended period is the median of the measurements retained by the selection. We intentionally do not compute an inverse-variance weighted mean, to avoid cases where a single, spuriously precise quoted error dominates the result. We adopt a conservative $1\sigma$ uncertainty, $\sigma_P = \max(\sigma_{\rm scatter},\,\tilde{\sigma}_{\rm quoted})$, where $\sigma_{\rm scatter}$ is the standard deviation of all the measurements within $20\%$ of the recommended period and $\tilde{\sigma}_{\rm quoted}$ the median uncertainty quoted by the sources behind that period. Where the quoted errors set the bar and are two-sided, the bar is split in the ratio of their median lower to their median upper side, with the mean of the two sides kept equal to $\sigma_P$; where the scatter sets it, the bar is symmetric.

Each star is finally ranked on a reliability tier, with the most reliable periods in
tier~1 and the lower tiers progressively admitting periods that carry a quality or physical
caveat; the tiers and the flags that drive them are defined in Appendix~\ref{app:gatetiers}. 
Ten of the $2716$ stars are ultimately excluded or left unresolved by the selection, leaving
$2706$ with a recommended period.

\subsection{Stellar mass}
\label{sec:mass}

The mass prior $\mathcal{M}_0\pm\sigma_{\mathcal{M}}$ entering the inference
(Sect.\,\ref{sec:inference}) is obtained from all the available mass estimates of each
star by the same single-value selection procedure used for the period. The mass sample is
both larger and less affected by discordance: $\sim$$32\,000$ measurements across $\sim$$4700$ stars,  
with inter-method offsets small against the star-to-star spread. The selection again adopts
the median of the chosen set with the same conservative error rule, with a few physically
motivated refinements (Appendix~\ref{app:mass}). 
The mass is recovered for essentially the entire selected sample, almost all of it at tier~1.

\subsection{Colour, effective temperature, and metallicity}
\label{sec:colour}

The colour $C_\star$ that enters the likelihood is the de-reddened {\it Gaia} DR3 $G_{\rm BP}-G_{\rm RP}$
index \citep{GaiaDR3}.
We subtract a line-of-sight reddening $E(G_{\rm BP}-G_{\rm RP})$ read from the three-dimensional extinction map of \citet{Vergely2022}, integrated to each star's adopted distance, the most precise of its estimates among {\it Gaia}~DR3, \citet{BailerJones2021}, TICv8 \citep{Stassun2019}, and \citet{Berger2020,Berger2026}, and propagate the reddening uncertainty into $\sigma_C$. The correction is small for most of this nearby sample and is detailed in Appendix~\ref{app:colour}.
 The effective temperature $T_{\rm eff}$ and metallicity
$\mathrm{[Fe/H]}$ are not used in the likelihood (the rotational model assumes solar
metallicity, Sect.\,\ref{sec:model}) but are selected with the same framework, for
classification and cross-checks. Here the selection follows a method hierarchy (interferometry for the temperature, where available, then spectroscopy and {\it Gaia}
GSP-Spec, then isochrone fitting, and finally photometry), with outlier rejection inside the
selected tier; the details are in Appendix~\ref{app:tefffehcolour}. The selected
temperature is spectroscopic for $\sim$$81\%$ of the sample and the metallicity for $\sim$$76\%$,
isochrone fitting supplying most of the remainder.

\subsection{The final sample}
\label{sec:finalsample}

A rotation period, a stellar mass, and a de-reddened colour, each with a declared uncertainty, are available together for $2497$ of the $2706$ stars with a recommended period.
Two physical conditions then define the sample on which the rotational model can be evaluated.
First, the model tracks span only $0.4$--$1.25\,\mathcal{M}_\odot$ in mass (Sect.\,\ref{sec:model}) and
$0.647$--$2.99$ in $C_\star$; the $239$ stars whose
selected mass \emph{or} de-reddened colour falls outside this domain are excluded. Second, a further $54$ stars whose
selected mass and radius
imply a surface gravity below its main sequence value ($\log g<4.0$)
are dropped, because gyrochronology is calibrated on main-sequence dwarfs alone, leaving $2204$
stars on which the rotational model is evaluated.
The stellar data catalogue is available in a machine-readable table\footnote{[Link to the
stellar catalogue deliverable to be inserted.]}. It lists the 4639 hosts for which at least one among rotation period, mass, effective temperature, metallicity, surface gravity and radius could be adopted.

%%%%%%%%%%%%%%%%%%%%%%%%%%%%%%%%%%%%%%%%%%%%%%%%%%%%%%%%%%%%%%

\section{Method}

\subsection{The gyrochronology model}
\label{sec:model}

Stellar ages are inferred by fitting an extended version of the rotational tracks constructed with the two-zone rotational evolution model of \citet{SL:26} (hereafter SL26). The model is applicable to stars in the mass range $0.4 \leq \mathcal{M}/\mathcal{M}_{\odot} \leq 1.25$ that possess a solar-like internal structure during the main sequence, consisting of a radiative core and a convective envelope. Conservatively, the requirement of a solar-like interior restricts the applicability of the model to main-sequence evolutionary phases. To ensure continuity and prevent hard-boundary effects, however, in this work we extend the rotational evolution tracks to the terminal-age main sequence (TAMS), as detailed below.

The free parameters of the model were calibrated using rotational data from a set of open clusters spanning ages from $\tau \simeq 0.1$ to $\simeq 4.0$ Gyr, where $\tau$ denotes stellar age. This calibration range is constrained both by the development of the slow-rotator sequence and by the availability of suitable observational data. The youngest cluster exhibiting a well-defined slow-rotator sequence is the Pleiades ($\tau \simeq 0.1$ Gyr), whereas the oldest open cluster with currently available rotational measurements is M67 ($\tau \simeq 4$ Gyr). The upper calibration limit also approximately coincides with the onset of the so-called weakened magnetic braking regime observed in solar-type stars \citep{2016Natur.529..181V,2017SoPh..292..126M,2025ApJ...986..120M}. While the mass dependence of this phenomenon remains under investigation, particularly toward lower stellar masses, it is not expected to affect the age range over which the present model is calibrated. The model neglects the effect of metallicity on the rotational evolution of solar-type stars.

   \begin{figure}[t!]
   \centering
   \includegraphics[width=\hsize]{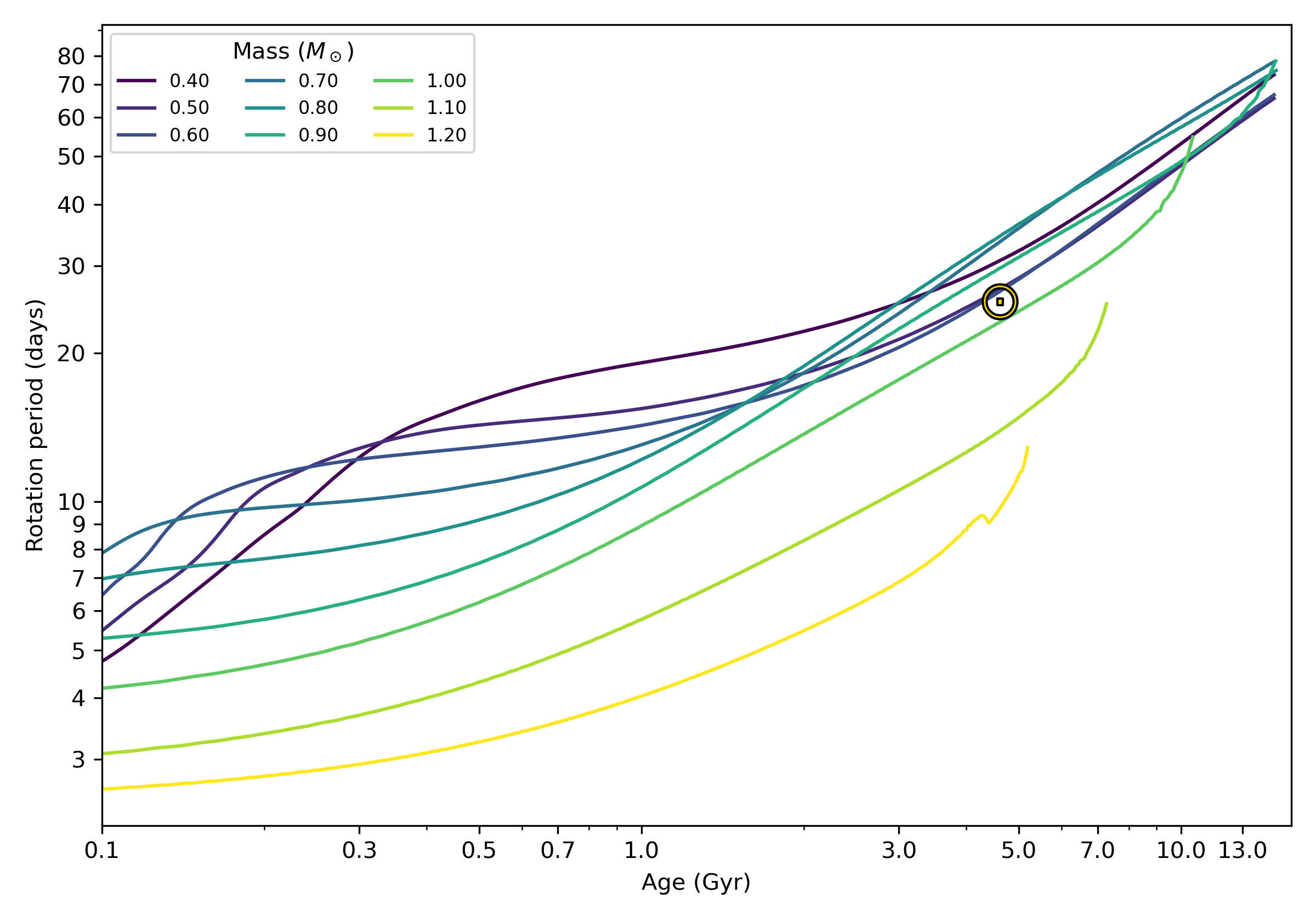}
      \caption{Rotational evolutionary tracks adopted in this work for representative stellar masses. The tracks are obtained by extrapolating the SL26 rotational evolution models up to the TAMS. The solar position is indicated by the solar symbol.}
         \label{fig:rotation_tracks}
   \end{figure}

In this work, the SL26 rotational tracks are extrapolated up to the terminal-age main sequence (TAMS), defined here as the age at which the central hydrogen abundance decreases below $X_{\rm c}<10^{-12}$. This extrapolation extends beyond both the age range used for model calibration and the regime in which weakened magnetic braking is expected to operate. Nevertheless, our primary objective is to place all stars in the sample on a homogeneous and continuous age scale. We therefore retain the extrapolated tracks at all ages, while treating ages above $\sim4$ Gyr as indicative. This limitation does not affect the main conclusions of this work, since the extrapolation is used primarily to provide a consistent age ordering for stars lying beyond the calibrated range. Furthermore, the planetary evolutionary phases of primary interest occur within the age interval over which the model has been calibrated (see Paper II).

The stellar structure quantities required by the model are obtained through interpolation of stellar evolutionary tracks computed with the Yale Rotational stellar Evolution Code (YREC) in its non-rotating configuration \citep{Demarque_etal:2008}. The colour index $C \equiv (G_{\rm BP}-G_{\rm RP})$ at each point along the rotational evolutionary tracks is computed from the corresponding values of $T_{\rm eff}$, $\log g$, and $\mathrm{[Fe/H]}$ using the YBC database of bolometric corrections\footnote{\url{https://sec.center/YBC/index.html}} \citep{2019A&A...632A.105C}. We adopted the Gaia DR3 photometric passbands \citep{2023A&A...674A...3M} and the PHOENIX spectral library \citep{2012EAS....57....3A}, assuming solar metallicity and negligible interstellar extinction.

Figure~\ref{fig:rotation_tracks} shows the rotational evolutionary tracks used in this work. For stars with $\mathcal{M}\gtrsim0.9\,\mathcal{M}_{\odot}$, the rotation period increases more rapidly as the TAMS is approached. According to the model, the decreasing angular velocity progressively reduces the wind-braking torque (see Eq.~2 of SL26), while the rapid expansion of the star causes its moment of inertia to increase. The latter effect produces additional spin-down and eventually dominates the evolution of the rotation period, even when angular-momentum loss through magnetic braking has become relatively inefficient. For a $1.2\,\mathcal{M}_{\odot}$ star, this departure from the earlier rotational evolution becomes noticeable at ages of $\sim5$ Gyr, whereas for a $1.0\,\mathcal{M}_{\odot}$ star it occurs at $\sim8$ Gyr. For stars with $\mathcal{M}\lesssim0.9\,\mathcal{M}_{\odot}$, the corresponding structural effect is expected only at ages exceeding the Hubble time.

   \begin{figure}[t!]
   \centering
   \includegraphics[width=\hsize]{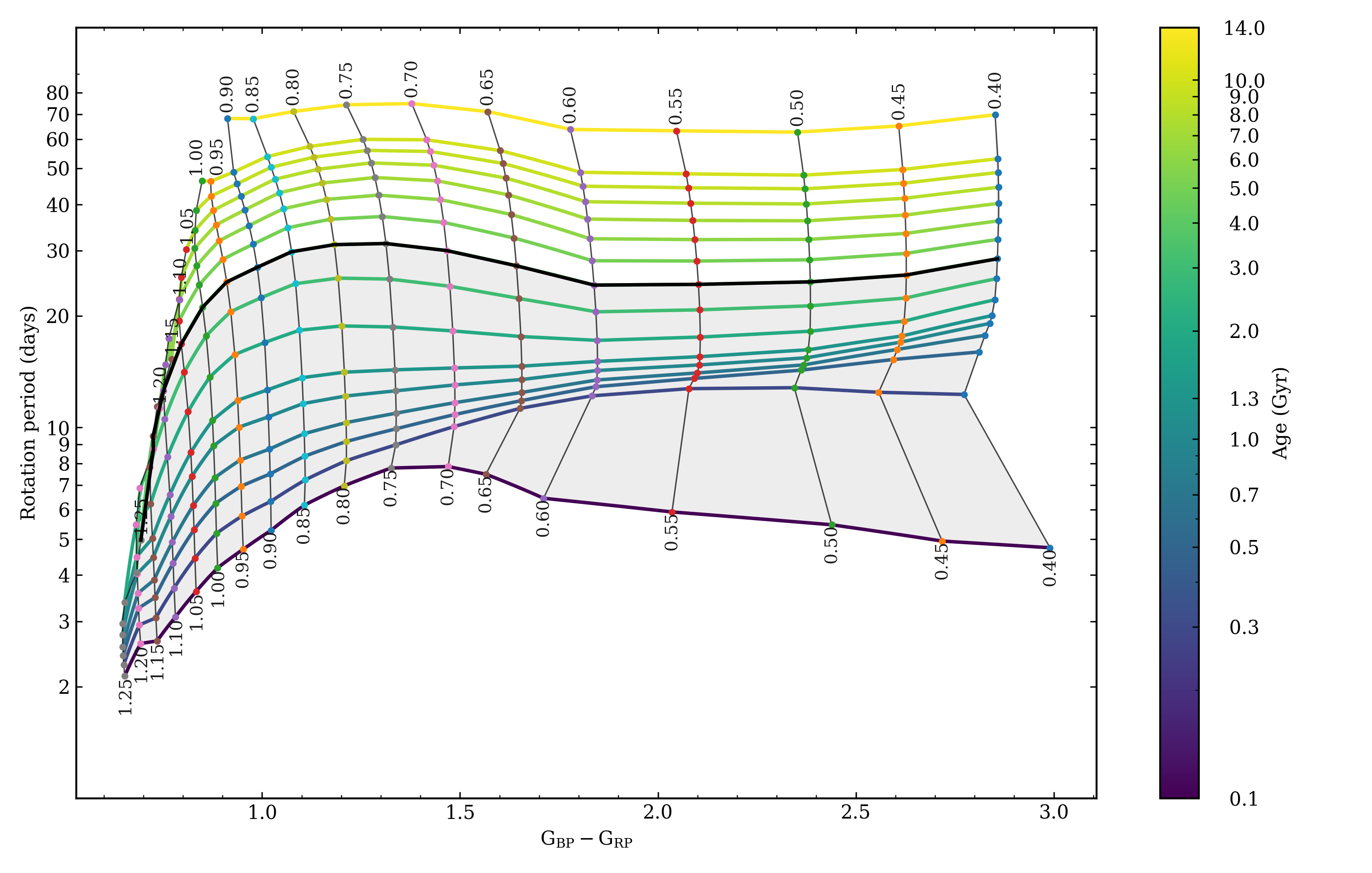}
      \caption{Rotational isochrones in the rotation period--Gaia $(G_{\rm BP}-G_{\rm RP})$ colour plane for the stellar models adopted in this work. Isochrones are colour-coded according to age, as indicated by the colour bar. Lines connecting points of equal initial stellar mass are shown in grey and labelled in units of $\mathcal{M}_{\odot}$. The thick black line marks the upper boundary of the age range over which the rotational model is calibrated using open-cluster data, and the shaded area indicates the corresponding calibrated region.}
         \label{fig:isochrones}
   \end{figure}

For $\tau \gtrsim 4$ Gyr, the model predicts that the convective envelope and radiative core are effectively coupled throughout the mass range considered here. Away from the structural changes associated with the TAMS, the subsequent rotational evolution approaches a Skumanich-like relation, $P\propto\tau^{1/2}$ \citep{2015A&A...584A..30L}, and the dependence of the rotational evolution on stellar mass becomes significantly weaker.

Figure~\ref{fig:isochrones} shows the resulting model isochrones in the rotation period--colour plane. Curves of constant age (coloured curves) are obtained by evaluating the rotational evolution model for stellar masses between $0.4$ and $1.25\,\mathcal{M}_{\odot}$, while curves of constant mass (nearly vertical curves) trace the evolutionary paths followed by stars of a given initial mass. The colour map identifies the corresponding ages (in Gyr), and the labels indicate initial stellar masses (in $\mathcal{M}_{\odot}$). The thick black curve marks the oldest rotational isochrone included in the calibration against open clusters ($\tau\simeq4$ Gyr), while the shaded region indicates the domain over which the model is directly constrained by observations. Isochrones at older ages are obtained by extrapolating the rotational evolution model up to the TAMS, as discussed above.

The geometry of the isochrone grid reflects the different regimes of rotational and structural evolution across the mass--age plane. During the phase in which angular momentum is transferred from the radiative core to the convective envelope, the surface spin-down is partially counteracted by the angular-momentum supply from the core, causing successive rotational isochrones to become more closely spaced. Once core and envelope approach rotational coupling, the isochrones become more regularly and widely spaced, as the subsequent evolution approaches the Skumanich-like dependence of rotation period on age. Owing to the strong mass dependence of the core--envelope coupling timescale, this crowding occurs at progressively younger ages with increasing stellar mass and is not apparent above approximately $0.75\,\mathcal{M}_{\odot}$ over the age range shown. At masses above approximately $1\,\mathcal{M}_{\odot}$, the geometry of both the isochrones and the constant-mass tracks is increasingly affected by stellar structural evolution toward the TAMS. These stars evolve towards redder colours and undergo additional spin-down as their radii and moments of inertia increase, even though magnetic braking becomes progressively less effective. The age sensitivity of the grid therefore does not vary monotonically with either age or mass, but reflects the interplay between core--envelope angular-momentum exchange, wind braking, and structural evolution.
 
The grid also illustrates the complementary information provided by the Gaia colour. Over most of the main sequence, the colour varies only weakly with age for a given stellar mass, so that the colour primarily constrains the stellar mass rather than the age. However, as stars approach the terminal-age main sequence, the evolutionary tracks bend toward redder colours owing to stellar expansion, causing tracks of different masses to converge. This behaviour introduces a partial degeneracy between stellar mass and age, which is mitigated in the Bayesian analysis by including independent literature estimates of the stellar mass as prior information.

\subsection{Inference of stellar ages from rotation periods and colours}
\label{sec:inference}

Stellar ages are inferred by comparing the observed surface rotation period and colour with predictions from the extended SL26 model grid described in Sect.\,\ref{sec:model}. 
For each combination of stellar mass and age, the model provides the corresponding rotation period ($P$) and Gaia colour index ($C$).

To obtain a continuous representation of the model predictions, we performed a linear interpolation over a Delaunay triangulation of the $(\mathcal{M},\tau)$ parameter space. This procedure yields the interpolated functions
$$
\hat{P}(\mathcal{M},\tau)
$$
and
$$
\hat{C}(\mathcal{M},\tau),
$$
which can be evaluated at any location within the convex hull of the model grid.

For a star with measured rotation period ($P_\star$) and colour ($C_\star$), we defined the likelihood function assuming independent Gaussian uncertainties,
$$
\ln \mathcal{L}
=
-\frac{1}{2}
\left[
\left(
\frac{\hat{P}(\mathcal{M},\tau)-P_\star}
{\sigma_P}
\right)^2
+
\left(
\frac{\hat{C}(\mathcal{M},\tau)-C_\star}
{\sigma_C}
\right)^2
\right].
$$
Here, $\sigma_P$ and $\sigma_C$ denote the uncertainties on the measured rotation period and colour, respectively. The uncertainty $\sigma_P$ includes both the observational uncertainty and an empirical estimate of the intrinsic width of the slow-rotator sequence, which are added in quadrature. Following \citet{2015A&A...584A..30L}, we adopt a fixed value of $\sigma = 0.6$\,d for this latter contribution. This term accounts for the observed dispersion of the slow-rotator sequence in open clusters and effectively captures the combined effects of intrinsic stellar variability, surface differential rotation, and residual model uncertainties.

Prior information on the stellar mass was incorporated through a Gaussian prior,
$$
p(\mathcal{M})
\propto
\exp\left[
-\frac{1}{2}
\left(
\frac{\mathcal{M}-\mathcal{M}_0}{\sigma_{\mathcal{M}}}
\right)^2
\right],
$$
where ($\mathcal{M}_0$) and ($\sigma_{\mathcal{M}}$) are the externally determined mass estimate and its uncertainty. 
A uniform prior was adopted for the stellar age over the range covered by the evolutionary models,
$$
p(\tau)=\mathrm{const}.
$$
We adopted a uniform prior in age because our goal is to derive age estimates that depend primarily on the observed stellar rotation and colour, while introducing the minimum possible amount of prior information. Since the sample is not drawn from a well-defined volume-limited stellar population and may be affected by complex selection effects, adopting a more informative age prior based on Galactic stellar population models could introduce biases that are difficult to quantify.

The posterior probability distribution is therefore
$$
p(\mathcal{M},\tau \mid P_\star,C_\star)
\propto
\mathcal{L}(P_\star,C_\star \mid \mathcal{M},\tau)\,
p(\mathcal{M})\,
p(\tau).
$$

The posterior was sampled using the affine-invariant ensemble Markov Chain Monte Carlo sampler implemented in the \texttt{emcee} package \citep{ForemanMackey2013}. We employed 32 walkers, each evolved for 10\,000 steps. The walkers were initialised around the prior mass estimate, and the first 2000 steps of each chain were discarded as burn-in.
The integrated autocorrelation time was approximately 50 steps, leaving
about 160 autocorrelation times in the post-burn-in portion of each chain
and ensuring adequate sampling of the posterior distributions.

The marginalised posterior probability distribution of the stellar age was then obtained by integrating over the nuisance parameter ($\mathcal{M}$),
$$
p(\tau \mid P_\star,C_\star)
=
\int
p(\mathcal{M},\tau \mid P_\star,C_\star)\,
d\mathcal{M}.
$$

The median of the marginalised age posterior was adopted as the nominal age estimate, while the 16th and 84th percentiles were used to define the corresponding ($1\sigma$) credible interval. This procedure naturally propagates the uncertainties in the observed rotation period, colour, and prior mass estimate into the final age determination.

%%%%%%%%%%%%%%%%%%%%%%%%%%%%%%%%%%%%%%%%%%%%%%%%%%%%%%%%%%%%%%

\section{Results}
\label{sec:results}

The results show that the rotational inference provides well-constrained ages for most stars with reliable period measurements. Of the 2204 stars for which an age estimate can be obtained, 1870 ($\simeq85\%$) have a bounded posterior distribution and are assigned a nominal age, whereas 240 ($\simeq11\%$) and 94 ($\simeq4\%$) yield lower and upper age limits, respectively. For stars with bounded posteriors, the inferred ages span approximately $0.15$--$21$ Gyr\footnote{The inferred age of 13 stars is larger than the Hubble time. The lowest confidence tier ($q_{\rm Age}=4$) is assigned to these. (See Appendix\,\ref{app:gatetiers})}, with a median of $3.59$ Gyr and an interquartile range of $2.19$--$5.21$ Gyr. The typical statistical uncertainty is relatively small, with median lower and upper uncertainties of about $0.40$ and $0.37$ Gyr, respectively, corresponding to a fractional uncertainty of approximately $12\%$. Ages extending beyond the open-cluster calibration range should nevertheless be interpreted with caution, as discussed in Sect.~\ref{sec:model}, because they rely on the extrapolated rotational tracks.

   \begin{figure*}[t!]
   \centering
   \includegraphics[width=\hsize]{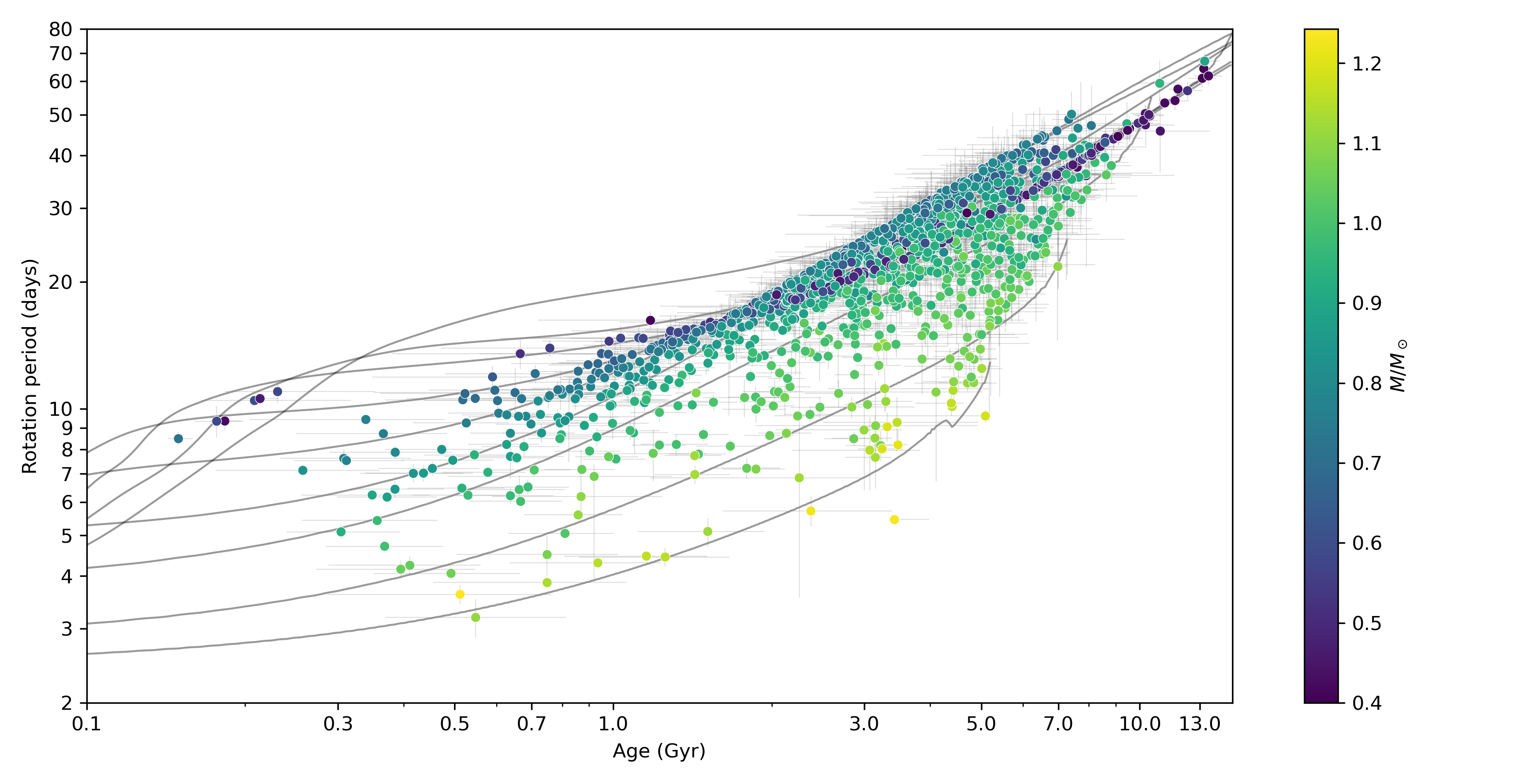}
      \caption{Rotation period as a function of inferred stellar age for stars with reliable rotation-period measurements (age tiers 1 and 2). Individual stars are colour-coded according to their inferred stellar mass, while the grey curves show the rotational evolutionary tracks of the extended SL26 model for different stellar masses. Grey error bars indicate the uncertainties on the inferred ages and measured rotation periods. The comparison illustrates the mass dependence of rotational evolution and the overall consistency between the inferred ages and the theoretical rotational tracks.}
         \label{fig:gyroages}
   \end{figure*}

Figure~\ref{fig:gyroages} shows the observed rotation periods as a function of the ages inferred with the Bayesian procedure described in Sect.~\ref{sec:inference}. The individual stars are colour-coded according to their inferred masses, while the grey curves represent the rotational evolutionary tracks predicted by the extended SL26 model. The resulting distribution displays a clear mass-dependent ordering, with stars of similar mass preferentially following the corresponding theoretical rotational sequences.

The overall distribution of the observations is consistent with the expected rotational evolution. At a given age, lower-mass stars generally occupy sequences characterised by different rotation periods from their higher-mass counterparts, reflecting the mass dependence of magnetic braking and of the internal angular-momentum evolution included in the model. Conversely, for a given mass, the rotation period generally increases with age as stars lose angular momentum through magnetised stellar winds. The observed stars broadly trace these theoretical evolutionary paths over the age interval where the sample is most densely populated.

The agreement is particularly evident in the main locus of the distribution, where the inferred stellar masses produce a smooth progression between neighbouring evolutionary tracks. This provides a useful consistency check on the simultaneous determination of mass and age: stars assigned different masses do not populate the period--age diagram randomly, but instead occupy the regions expected from the corresponding mass-dependent rotational evolution. Some dispersion around the individual tracks is nevertheless present, as expected from observational uncertainties in the rotation periods and ages, surface differential rotation, the intrinsic width of the slow-rotator sequence, and residual limitations of the rotational evolution model.

At the oldest inferred ages, the interpretation requires additional caution because the rotational tracks extend beyond the age range directly calibrated with open-cluster data. In this regime, the inferred ages depend on the extrapolation of the SL26 model discussed in Sect.~\ref{sec:model} and should therefore be regarded as indicative. The period--age diagram nevertheless provides a homogeneous representation of the rotational evolution of the full sample and shows that the inferred stellar parameters preserve the expected dependence of rotational evolution on stellar mass.

   \begin{figure*}[t!]
   \centering
   \includegraphics[width=0.8\hsize]{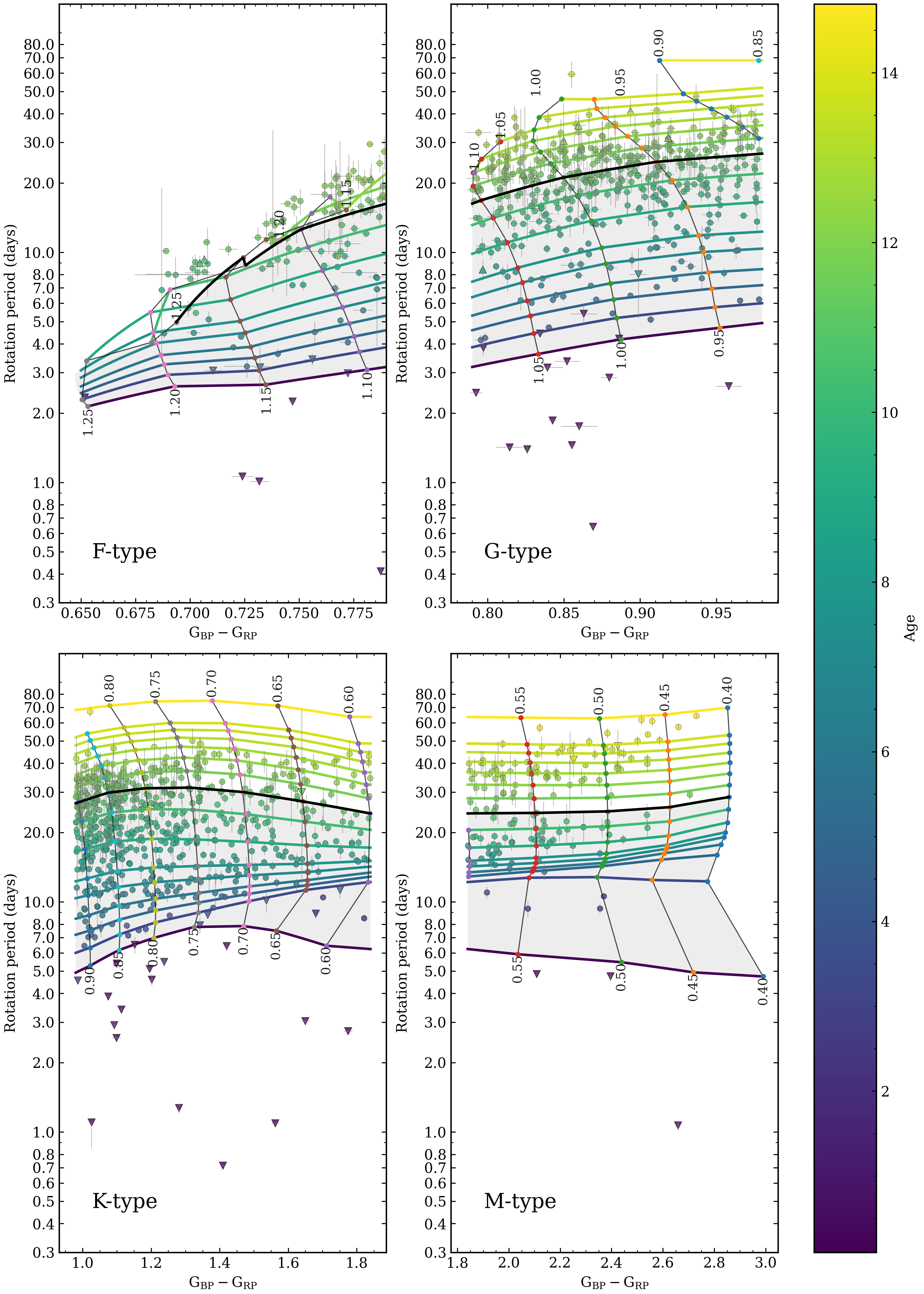}
      \caption{Observed rotation periods as a function of de-reddened Gaia $(G_{\rm BP}-G_{\rm RP})$ colour for the F-, G-, K-, and M-type stars for age tiers 1 and 2. Coloured curves show the rotational isochrones of the extended SL26 model, with colour indicating age as given by the colour bar. Grey curves connect points of equal stellar mass, with the corresponding masses (in $\mathcal{M}_{\odot}$) indicated by the labels. The thick black curve marks the oldest isochrone within the age range directly calibrated with open clusters, and the grey shaded region indicates the corresponding calibrated domain. Individual stars are shown with their observational uncertainties and are colour-coded according to their inferred ages. Downward triangles denote objects for which only an upper age limit could be inferred. The four panels illustrate the different mass and age dependence of rotational evolution across the F-, G-, K-, and M-star regimes.}
         \label{fig:isochrones_stype}
   \end{figure*}

Figure~\ref{fig:isochrones_stype} compares the observed rotation periods of the stars with reliable photometric rotation measurements to the rotational isochrones adopted in this work. The coloured curves represent lines of constant age in the period--colour plane, while the thick black line marks the approximate upper limit of the age range over which the rotational model has been calibrated using open clusters. The shaded region identifies the corresponding domain of validity of the calibration. Symbols are colour-coded according to the gyrochronological age inferred from the Bayesian analysis. Triangles indicate stars for which the posterior probability distribution reaches one of the boundaries of the model grid, so that only an upper or lower limit on the age can be derived.

Overall, the distribution of the data follows the expected morphology of the rotational isochrones. The majority of stars populate the calibrated region of the diagram and lie close to the model predictions, indicating that the extended SL26 model provides a satisfactory description of the observed period--colour relation over the mass range considered. As expected from magnetic braking, stars with longer rotation periods are preferentially associated with older inferred ages, although the age sensitivity decreases toward lower stellar masses, where rotational evolution proceeds more slowly and the isochrones become increasingly crowded.

The comparison also highlights the different behaviour across spectral types. F- and G-type stars exhibit a broad distribution of rotation periods spanning most of the calibrated age range, allowing relatively precise age estimates. In contrast, the K- and especially M-type stars are concentrated in regions where the rotational isochrones are more closely spaced. In these cases, a given uncertainty in the measured rotation period translates into a larger uncertainty in the inferred age, reflecting the intrinsically weaker age dependence of stellar rotation at lower masses.

A fraction of the sample lies outside the calibrated region. Most of these objects are slow rotators located above the oldest calibrated isochrone, for which the Bayesian inference returns lower limits on the stellar age. Conversely, a smaller number of rapidly rotating stars fall below the youngest calibrated isochrone and are assigned upper age limits. These stars are retained in the analysis to preserve a homogeneous ordering of the sample in terms of rotational age, but their inferred ages should be regarded as indicative, since they rely on an extrapolation of the rotational model beyond the range constrained by open-cluster observations.

An additional consistency check is provided by comparing the stellar parameters predicted by the evolutionary model at the inferred mass and age with independent values collected from the literature. The inferred masses show very good agreement with the literature estimates. For the 1870 stars with bounded age posteriors, the median difference, defined as $\mathcal{M}_{\rm model}-\mathcal{M}_{\rm lit}$, is only $0.031\,\mathcal{M}_{\odot}$, with a robust scatter of $\simeq0.037\,\mathcal{M}_{\odot}$. In relative terms, the median offset is about $3.7\%$, while the median absolute deviation between model and literature masses is approximately $4.2\%$. No significant trend of the mass residuals with stellar mass is apparent. This agreement is expected to some extent because the literature mass estimates enter the inference as Gaussian priors; nevertheless, it demonstrates that the posterior solutions remain consistent with the independent mass constraints rather than being driven toward substantially different regions of the evolutionary grid.

The effective temperatures predicted by the model provide a more independent validation of the inferred stellar parameters. The model temperatures reproduce the literature values with a modest systematic offset: $T_{\rm eff,model}-T_{\rm eff,lit}$ has a median of approximately $+88$ K and a robust scatter of about $116$ K. The median absolute difference is $\simeq112$ K, corresponding to only $\sim2.2\%$ in relative terms. Thus, despite $T_{\rm eff}$ not being used directly as an observational constraint in the gyrochronological inference, the combination of inferred mass and age, together with the adopted stellar evolutionary tracks and colour transformations, yields effective temperatures that are in good overall agreement with independent literature determinations. The small positive temperature offset may reflect residual systematics in the adopted stellar models, bolometric corrections, or the heterogeneous temperature scales of the literature compilation, but it is sufficiently small not to indicate a significant inconsistency in the inferred stellar parameters.

Taken together, the agreement in both mass and effective temperature supports the internal consistency of the gyrochronological solutions. In particular, the comparison with $T_{\rm eff}$ provides a useful external check that the inferred locations of the stars in the mass--age plane are compatible with their independently determined atmospheric properties, while the distribution of age limits identifies those objects for which the rotation period lies outside, or close to the boundaries of, the region most strongly constrained by the calibrated rotational models.

   \begin{figure*}[t!]
   \centering
   \includegraphics[width=\hsize]{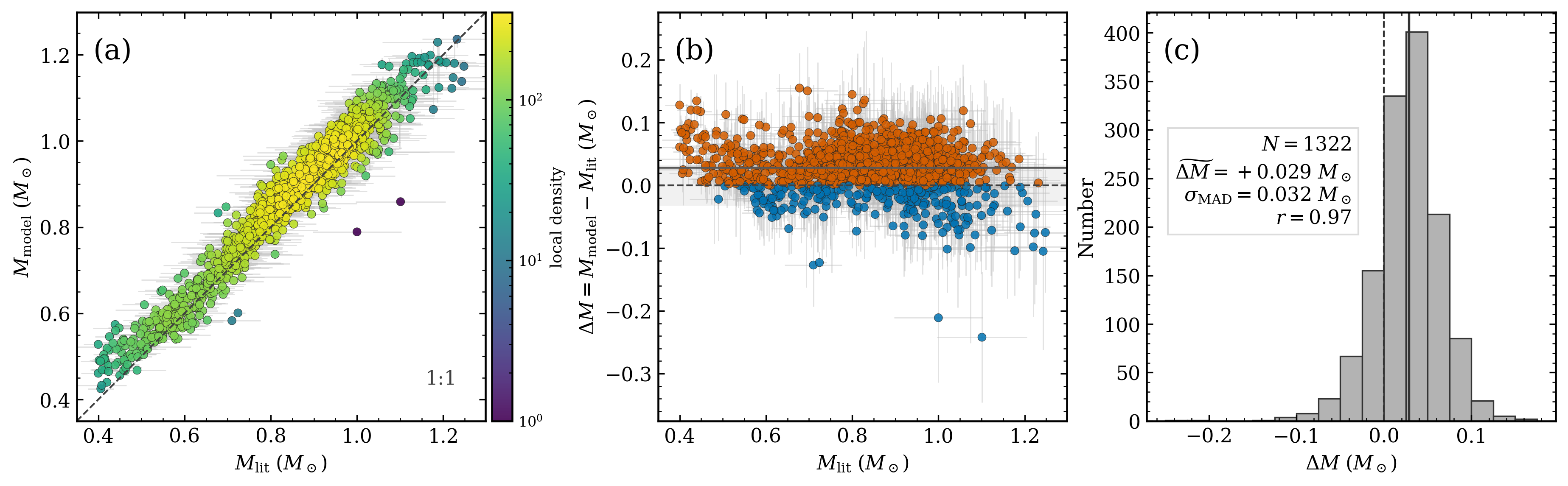}
   \includegraphics[width=\hsize]{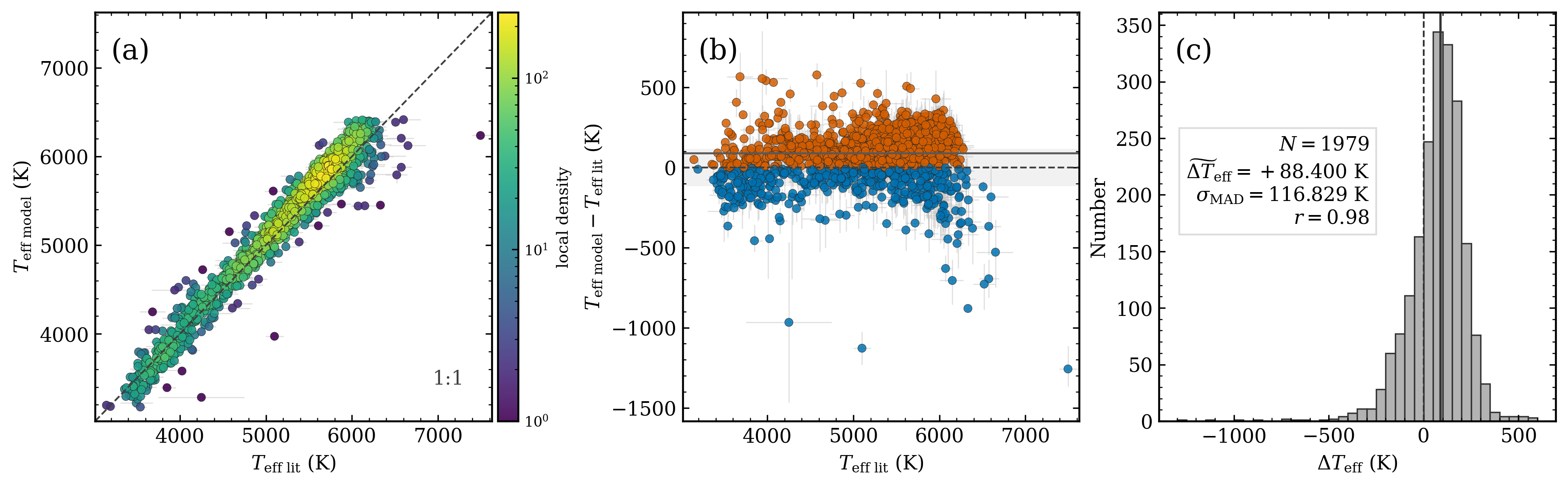}
      \caption{Comparison between the stellar parameters inferred from the gyrochronological analysis and the corresponding literature values. \textit{Top row:} stellar mass. \textit{Bottom row:} effective temperature. In each row, panel (a) compares the model-inferred and literature values; the dashed line indicates the one-to-one relation, and points are colour-coded according to the inferred stellar age. Panel (b) shows the residuals, defined as the model-inferred minus literature value, as a function of the literature value. Orange and blue symbols indicate positive and negative residuals, respectively; the solid line marks the median residual and the shaded band its robust dispersion. Panel (c) shows the distribution of the residuals; the solid and dashed vertical lines indicate zero and the median residual, respectively. The inset reports the number of stars, the median residual, the median absolute deviation (MAD) converted to a robust standard deviation, and the Pearson correlation coefficient $r$. For the stellar masses, the median difference is $\widetilde{\Delta M}=+0.029\,\mathcal{M}_{\odot}$, with $\sigma_{\rm MAD}=0.032\,\mathcal{M}_{\odot}$ and $r=0.97$. For the effective temperatures, $\widetilde{\Delta T_{\rm eff}}=+88.4$ K, $\sigma_{\rm MAD}=116.8$ K, and $r=0.98$. The agreement in mass is not an independent validation because the literature mass enters the inference as a prior, whereas $T_{\rm eff}$ is not used directly as an observational constraint and therefore provides an additional consistency check on the inferred stellar parameters.}
         \label{fig:MassTeffComparison}
   \end{figure*}

\begin{table*}
\caption{Adopted stellar parameters for the exoplanet-host stars considered
in this work. A portion of the table is shown here for guidance regarding
its form and content. The complete table, including uncertainties,
provenance information, reliability tiers, and quality flags, is available
at the CDS.}
\label{tab:stars}
\centering
\begin{tabular}{lrrrrrr}
\hline\hline
Gaia DR3 source ID &
$P_{\rm rot}$ &
$\mathcal{M}$ &
$(G_{\rm BP}-G_{\rm RP})$ &
$T_{\rm eff}$ &
$\mathrm{[Fe/H]}$ &
$\log g$ \\
&
(d) &
($\mathcal{M}_{\odot}$) &
(mag) &
(K) &
(dex) &
(dex) \\
\hline
1986772848034988544 & 10.456 & 0.6  & 1.702 & 4287 &&      \\
704967037090946688  & 39.00  & 0.92 & 1.006 & 5242 & 0.33 &  4.47 \\   \ldots & \ldots & \ldots & \ldots & \ldots & \ldots & \ldots \\
\hline
\end{tabular}
\tablefoot{Only a subset of the columns is shown. The complete table is
available in electronic form at the CDS.}
\end{table*}

The stellar parameters and gyrochronological results used in this work are provided as three electronic tables available at the CDS. Table~\ref{tab:stars} contains the adopted stellar parameters for the 4639 confirmed exoplanet-host stars considered in this study. The selection criteria of the stellar parameter are described in Appendix\,\ref{app:recon}. In addition to the source identifiers and Gaia DR3 astrometry, it provides the adopted rotation period, stellar mass, de-reddened Gaia $(G_{\rm BP}-G_{\rm RP})$ colour, effective temperature, metallicity, surface gravity, and stellar radius, together with their uncertainties. The table also reports the provenance and reliability tier of the adopted quantities, as well as flags and caveats relevant to the rotation-period measurements and to the consistency of the stellar parameters.

\begin{table*}
\caption{Gyrochronological ages and model stellar parameters. A portion of
the table is shown here for guidance regarding its form and content. The
complete table is available at the CDS.}
\label{tab:ages}
\centering
\begin{tabular}{lrrrrrrr}
\hline\hline
Gaia DR3 source ID &
$\tau$ &
$\tau_{\rm low}$ &
$\tau_{\rm upp}$ &
$P_{\rm model}$ &
$\mathcal{M}_{\rm model}$ &
$T_{\rm eff,model}$ &
$\log g_{\rm model}$ \\
&
(Gyr) &
(Gyr) &
(Gyr) &
(d) &
($\mathcal{M}_{\odot}$) &
(K) &
(dex) \\
\hline
1986772848034988544 & 0.208 & 0.165 & 0.317 & 11.025 & 0.633 & 4102 & 4.709 \\
704967037090946688  & 6.477 & 6.213 & 6.75. & 38.98 & 0.874 & 5305 & 4.525 \\
% Insert a few representative rows from ages.dat here
\ldots & \ldots & \ldots & \ldots & \ldots & \ldots & \ldots & \ldots \\
\hline
\end{tabular}
\tablefoot{The reported age corresponds to the median of the marginalised
age posterior, with the lower and upper bounds derived from the 16th and
84th percentiles, respectively. For sources for which only an upper or
lower age limit can be determined, the corresponding limit is identified
in the complete electronic table.}
\end{table*}

Table~\ref{tab:ages} contains the results of the gyrochronological analysis for the 2204 stars for which an age could be inferred. The reported age corresponds to the median of the marginal posterior distribution, with the 16th and 84th percentiles defining its uncertainty interval. When the posterior reaches a boundary of the model grid, or when the observed rotation period lies beyond the grid, the result is instead reported as an upper or lower age limit. A reliability tier is provided for each age and reflects the reliability of the observational quantities entering the inference. The table also gives the rotation period, stellar mass, de-reddened colour, effective temperature, and surface gravity predicted by the gyrochronological model at the inferred posterior solution. These quantities allow a direct comparison between the adopted stellar properties in Table~\ref{tab:stars} and the corresponding model predictions. We note that the model mass is not an independent validation of the inference because the literature mass enters as a prior, while the comparison of the model effective temperature with the literature value provides a useful, albeit model-dependent, consistency check.

The bibliographic references associated with the catalogue are supplied as an additional table at CDS.

%%%%%%%%%%%%%%%%%%%%%%%%%%%%%%%%%%%%%%%%%%%%%%%%%%%%%%%%%%%%%%

\section{Conclusions}
\label{sec:conclusions}

This work presents a homogeneous determination of gyrochronological ages for a large sample of confirmed exoplanet-host stars and constitutes the first paper in the \textit{Tracking the evolution of sub-Neptunes} series. The primary aim of this first study is to establish the stellar-age framework required to follow the evolution of close-in planetary systems on Gyr timescales. To this end, we combined reliable stellar rotation measurements with the two-zone rotational evolution model of \citet{SL:26}, constructing a homogeneous age scale for main-sequence stars in the mass range $0.4$--$1.25\,\mathcal{M}_{\odot}$.

The SL26 model is calibrated against the slow-rotator sequences of open clusters from approximately $0.1$ to $4$ Gyr. For the purposes of this work, we extended the rotational evolutionary tracks up to the TAMS, allowing suitable stars outside the calibrated age interval to be placed on the same continuous rotational-age scale. These extrapolated ages should be regarded as indicative, particularly at old ages where weakened magnetic braking may modify the rotational evolution. Importantly, this extrapolation does not affect the empirical calibration of the younger stars that constitute the regime of primary interest for tracing planetary evolution.

Stellar ages were inferred within a Bayesian framework by simultaneously comparing the observed rotation period and de-reddened Gaia $(G_{\rm BP}-G_{\rm RP})$ colour with the model predictions. Independent stellar masses from the literature were incorporated as Gaussian priors, reducing the mass--age degeneracy that becomes increasingly important toward the TAMS. The likelihood accounts both for observational uncertainties in the rotation period and for the empirical width of the slow-rotator sequence following \citet{2015A&A...584A..30L}. The resulting posterior distributions therefore propagate the relevant observational and model uncertainties and naturally identify objects for which the available rotational information provides only an upper or lower age limit.

From a catalogue of 4639 confirmed exoplanet-host stars, we obtained gyrochronological constraints for 2204 objects with suitable stellar parameters and reliable rotation measurements. Their distribution in the period--age plane follows the expected mass-dependent rotational evolutionary sequences, providing a coherent ordering of the sample from young, rapidly rotating stars to old, slowly rotating systems. Objects outside the region directly calibrated by open clusters are explicitly identified, allowing the user to distinguish robust gyrochronological ages from estimates that depend on extrapolation of the rotational model.

A major advantage of this approach is the precision that can be achieved for main-sequence stars. Stellar ages are notoriously difficult to determine on the main sequence because many of the classical age diagnostics evolve only weakly during this phase. Isochrone fitting, in particular, generally provides limited age discrimination for unevolved low-mass stars, while other methods often rely on activity indicators or heterogeneous empirical calibrations. By exploiting the strong and systematic evolution of stellar rotation, gyrochronology provides substantially greater age resolution over the domain in which rotational evolution is empirically calibrated. For the stars with well-constrained posterior distributions in our sample, the typical formal age precision reaches the $\sim10$--$15\%$ level. The combination of this precision with a homogeneous analysis of a large sample of planet-hosting main-sequence stars represents, to our knowledge, an unprecedented age data set for studying the temporal evolution of close-in planetary systems.

The inferred stellar properties provide additional checks on the consistency of the analysis. Posterior stellar masses remain in close agreement with the independent literature estimates used as priors. More significantly, the effective temperatures predicted by the stellar evolutionary models at the inferred masses and ages reproduce the literature $T_{\rm eff}$ values well, even though effective temperature is not directly included as an observational constraint in the gyrochronological inference. The latter comparison therefore provides an independent, albeit model-dependent, consistency test of the inferred locations of the stars in the mass--age plane.

The homogeneous age scale established here opens the possibility of investigating planetary evolution directly as a function of time rather than by comparing heterogeneous stellar populations. This is particularly important for sub-Neptunes, whose observed properties retain the imprint of several evolutionary processes operating from the first few hundred Myr to Gyr timescales. The subsequent papers in the \textit{Tracking the evolution of sub-Neptunes} series will exploit the ages derived here to follow the evolution of the sub-Neptune population and its distribution in planetary parameter space, with particular emphasis on the emergence and evolution of the radius valley. The combination of precise stellar ages with the rapidly growing sample of well-characterised exoplanets provides an opportunity to place direct observational constraints on the timescales governing the evolution of close-in planets.

The adopted stellar parameters, their literature provenance, and the gyrochronological results are made available through the CDS in three interconnected tables. Besides providing the age catalogue underlying this series, these data preserve the information required to assess the reliability of individual estimates and to update the analysis as improved rotation periods and stellar parameters become available.

%%%%%%%%%%%%%%%%%%%%%%%%%%%%%%%%%%%%%%%%%%%%%%%%%%%%%%%%%%%%%%

%%%%%%%%%%%%%%%%%%%%%%%%%%%%%%%%%%%%%%%%%%%%%%%%%%%%%%%%%%%%%%

%%%%%%%%%%%%%%%%%%%%%%%%%%%%%%%%%%%%%%%%%%%%%%%%%%%%%%%%%%%%%%
\begin{acknowledgements}
This work was partially supported by the European Union -- NextGenerationEU, under the Italian Ministry of University and Research (MUR) National Recovery and Resilience Plan (PNRR), Mission 4, Component 2, Investment 1.1, PRIN 2022, project No. 2022HY2NSX, ``CHRONOS: adjusting the clock(s) to unveil the CHRONO-chemo-dynamical Structure of the Galaxy'' and the European Union's Horizon 2020 research and innovation programme under Grant Agreement No. 824135 (SOLARNET -- Integrating High Resolution Solar Physics).
\end{acknowledgements}

\bibliographystyle{aa} % style aa.bst
\bibliography{GyroAgeAppPaper.bib} % your references Yourfile.bib

\begin{appendix}

\nolinenumbers

%%%%%%%%%%%%%%%%%%%%%%%%%%%%%%%%%%%%%%%%%%%%%%%%%%%%%%%%%%%%%%

\section{Selection of the stellar parameters}\label{app:recon}

This appendix gives the details behind the selection of the rotation period, stellar mass, effective temperature, metallicity, surface gravity, and radius, in pipeline order: the full decision tree, the discriminating criteria and their physical or statistical motivation.
We describe here only the selection \emph{algorithm}; the integrated host-star database it operates on, with the full per-source bibliography, is presented in the companion paper (Paper~II).

\subsection{Design principles and outcome classification}
\label{app:design}

The reduction to a single value is a \emph{selection}, not an average: the candidates can be physically distinct (e.g., harmonic families or non-rotational contaminants, in the case of period measurements; Sect.~\ref{sec:rotation}). 
As a general rule, sets containing three or more compatible measurements are represented by their median rather than by an inverse-variance weighted mean. Specific two-measurement branches may instead adopt an unweighted or source-weighted mean, as detailed below. First, the central value is the \emph{median} of the chosen set, never an inverse-variance weighted mean: the literature errors are mutually inconsistent, so a single source with a drastically underestimated error would otherwise dominate. Second, every output error follows the same conservative prescription, the larger of the scatter of the measurements behind the value and the uncertainties quoted by the sources behind it,
\begin{equation}
\sigma \;=\; \max\!\,\,\bigl(\,\mathrm{std}\{\text{set}\},\ \mathrm{median}\{\text{quoted errors}\}\,\bigr),
\label{eq:comb_error}
\end{equation}
so that the bar cannot fall below the inter-source scatter. 
%%%%%
The error is evaluated on a different set for the rotation period than for the other parameters. For the period it is the \emph{support set} of the adopted value, the measurements that agree with it within $20\%$: the standard deviation is taken over the whole window, so that a measurement discarded by the selection can still widen the bar, while the median uncertainty comes only from the measurements that set the value. For the other parameters both terms are evaluated on the measurements retained by the selection. This support window should not be confused with the generally narrower consensus tolerances used by the parameter-selection algorithms described below.
Where the uncertainties quoted by the sources that set the value are two-sided, the bar is split in the ratio of their median lower to their median upper side, with the mean of the two sides kept equal to $\sigma$; otherwise it is symmetric.
%%%%%
Third, an ambiguous case is passed to the last-resort criteria of Appendix~\ref{app:levers}, and when these also fail no value is adopted and the star is released without a period. 
Finally, within each branch the rules form a strictly \emph{ordered sequence}, stopping at the first that resolves the case.

\subsection{Pre-filtering and physical limits}
\label{app:guards}

\subsubsection{Removing non-detections and duplicates}
\label{app:guards:clean}

Three deterministic cleaning steps run before any value is selected. First, measurements flagged as $v\sin i$-derived or as upper limits are discarded, since a projected velocity ($\sin i\le1$) only bounds the period from above; the same applies to values a source only assumes, to periods predicted from an activity--rotation relation, and to values the source itself reports as spurious.
Second, a value republished verbatim by several catalogues (identical number and error bars) is counted once.
Third, compilation catalogues that mix their own measurements with re-quotes of earlier work (\citealt{Gaidos2023,Bowler2023,David2021}) contribute a period only through the measurement they endorse, so a re-quote duplicating an already-present source is discarded.

\subsubsection{Excluding non-rotational periods}
\label{app:guards:vetoes}

Some catalogue ``periods'' are not stellar rotation at all. Two classes are identified by physical limits derived from first principles, a third by a documented literature decision.

\paragraph{The break-up (critical-rotation) period.}
A star cannot rotate faster than its centrifugal break-up limit. Following the Roche model of \citet{Townsend2004}, in which the central condensation is treated as a point mass and the equatorial radius $R_{\rm e}$ inflates to $R_{\rm e} = \tfrac{3}{2} R_{\rm p}$ at critical rotation (where $R_{\rm p}$ is the polar radius),\footnote{The break-up period is the equatorial rotation period at the critical velocity, $P_{\rm c}=2\pi R_{\rm e}/v_{\rm c}$, with $v_{\rm c} = \sqrt{G\mathcal{M}/R_{\rm e}} = \sqrt{2G\mathcal{M}/3R_{\rm p}}$ \citep[][their Eq.~1]{Townsend2004}; taking the catalogue radius as the polar radius $R_{\rm p}$ (valid for the slow rotators of interest) gives the constant of Eq.~(\ref{eq:pc}), $2\pi(3/2)^{3/2}\sqrt{R_\odot^3/G\mathcal{M}_\odot}=0.213$\,d.
%, i.e.\ $1.84\times$ the naive point-mass Keplerian value $0.116\,(R/R_\odot)^{3/2}(\mathcal{M}/\mathcal{M}_\odot)^{-1/2}$\,d
} the corresponding critical period is
\begin{equation}
\label{eq:pc}
P_{\rm c} = 0.213\,\left(\frac{R}{R_\odot}\right)^{3/2}
\left(\frac{\mathcal{M}}{\mathcal{M}_\odot}\right)^{-1/2}\ {\rm d}.
\end{equation}
Here $R$ and $\mathcal{M}$ are the \emph{selected} radius and mass (Appendix~\ref{app:mass}).
Any measurement with $P < P_{\rm c}$ is rotationally impossible and is interpreted as a non-rotational periodicity contaminating the catalogue, and so removed.

\paragraph{The projected-velocity ceiling.}
The projected rotational velocity sets a hard upper limit on the period, because $\sin i \le 1$:
\begin{equation}
\label{eq:pmax}
P_{\rm max} = 50.6\,\frac{R/R_\odot}{v\sin i\,[{\rm km\,s^{-1}}]}\ {\rm d},
\end{equation}
the numerical constant being $2\pi R_\odot/(86400\,{\rm s}\times 1\,{\rm km\,s^{-1}})$, again with the selected $R$ and, where more than one source measures it, the median $v\sin i$. A candidate with $P > 1.3\,P_{\rm max}$ is discarded in the decision tree (Appendix~\ref{app:tree}), unless every candidate of that star exceeds it. The factor $1.3$ absorbs the joint uncertainty in $R$ and $v\sin i$ (typically $\sim$$4\%$ and $\sim$$8\%$ above the floor at which the ceiling is applied, $\sim$$9\%$ combined) so that the constraint is enacted only on unambiguous violations. The ceiling is reliable only for sufficiently rapid rotators.  
It is therefore applied only when $v\sin i$ is at least $4\,{\rm km\,s^{-1}}$ and is not itself an upper limit.
Below this floor the spectroscopic line broadening is degenerate with macroturbulence: the reported $v\sin i$ is then an overestimate and $P_{\rm max}$ collapses to a spuriously small value.

\paragraph{Documented literature exclusions.}
A small set of stars are excluded by a documented literature decision, because their catalogue ``periods'' are physically not rotational. Among these are three Herbig\,Ae/Be disc-variability A-stars (AB\,Aur, HD\,97048, MWC\,758), whose photometry traces the circumstellar disc rather than the stellar surface. By contrast, the cool young hosts (the T~Tauri stars V1298\,Tau, AU\,Mic, PDS\,70, V830\,Tau) show genuine spot-modulation rotation and are left in the sample.

\subsection{The discriminating criteria}
\label{app:levers}

The criteria invoked throughout the decision tree of Appendix~\ref{app:tree} are collected here, grouped by function.

\subsubsection{Source weights and reliability}
The weight assigned to each catalogue and the division of the measurements into credible and low-reliability run through the whole decision tree of Appendix~\ref{app:tree}. The weight sets the mean of a near-concordant pair, scores the clusters, and breaks the ties that no other rule resolves. A low-reliability measurement loses to a credible one wherever the two meet.

\paragraph{Empirical source weights.}
Each catalogue is assigned a scalar weight by a two-pass procedure. In the first pass an unweighted reference period $P_{\rm ref}$ is constructed for each star with three or more measurements, every catalogue counting equally. When no measurement lies more than $25\%$ from the median, the reference is the median itself. When some measurements do, they are discarded and the reference is the median of the remaining values, provided that the survivors constitute a majority and agree within $15\%$. When even this criterion fails, the measurements are split into clusters wherever two consecutive periods differ by more than a factor $1.30$, and the reference is the median of the most populated cluster among those whose median lies below $1.3\,P_{\max}$ (Eq.~\ref{eq:pmax}).

The same first-pass reference periods are used to characterise the empirical behaviour of each measurement method. For a method $m$, we define the harmonic-ratio rates
\begin{equation}
p_k(m) \equiv
P\!\left(P_{\rm meas}/P_{\rm ref}\simeq k\,\middle|\,m\right),
\label{eq:methodrates}
\end{equation}
where $k\in\{1/3,1/2,1,2,3\}$ and agreement with a given ratio is defined within $15\%$. Thus, $p_1(m)$ is the empirical agreement rate of method $m$, while $p_{1/2}(m)$ and $p_2(m)$ are its halving and doubling rates, respectively; $p_{1/3}(m)$ and $p_3(m)$ describe the corresponding third- and triple-period aliases; the five rates do not sum to unity, the remainder being the measurements that stand in no harmonic relation to the reference. These rates are measured by pooling all measurements of the stars that have a first-pass reference, obtained with the same method across the catalogues in which that method is used, and constitute an \emph{internal calibration} of the measurement methods: they describe their empirical behaviour within the present compilation rather than externally calibrated probabilities of recovering the true stellar rotation period.

In the second pass every source's measurements are classified as agreeing with the first-pass reference within $15\%$ or not. The source weight is then defined as the shrinkage estimate
\begin{equation}
%w({\rm source}) = \frac{n_{\rm agree} + \alpha\,p_{\rm ref}({\rm source})}{N_{\rm source} + \alpha}\,,
w(s) = \frac{n_{\rm agree}(s)+\alpha\,p_1[m(s)]}{N_s+\alpha}\,,
\label{eq:srcweight}
\end{equation}
where $n_{\rm agree}$ and $N_s$ are the number of agreeing measurements and total number of measurements from source $s$, respectively, 
%and $p_{\rm ref}({\rm source})$ is a per-technique agreement prior evaluated at the technique used by the source (its most frequent one, for the few catalogues that use several): the fraction of all measurements made with that technique, pooled across every catalogue that uses it, that fall within $15\%$ of the first-pass reference. 
and $m(s)$ is the method used by that source (or its most frequent method for the few catalogues employing more than one).
%For the present compilation these priors run from $\approx0.95$ for the pipeline of \citet{Santos2019,Santos2021}, through $\approx0.87$ for bare auto-correlation, down to $0.37$ for the machine-learning periods of \citet{Kamai2025}; $\alpha=10$ pulls small-sample sources towards this prior.
The parameter $\alpha=10$ shrinks the empirical agreement rate of sources with small samples towards the corresponding method-level rate $p_1[m(s)]$.

For the present compilation, $p_1(m)$ ranges from $\approx0.95$ for the pipeline of \citet{Santos2019,Santos2021}, through $\approx0.87$ for bare auto-correlation, down to $0.37$ for the machine-learning periods of \citet{Kamai2025}. Using an unweighted reference in the first pass ensures that the reference periods, and hence the agreement counts, are not themselves determined by the source weights. The method-level rates are less independent when a method is represented by a single catalogue, in which case $p_1(m)$ reduces to that catalogue's own agreement rate. Among the catalogues contributing most of the measurements, the resulting source weights range from $\approx0.99$ for the auto-correlation pipeline of \citet{McQuillan2013} to $\approx0.37$ for \citet{Kamai2025}. A source weight discriminates between two candidate periods only when the corresponding weights differ by at least $0.10$.

Thus, $w(s)$ measures the overall empirical reliability of a \emph{source}, whereas the set of rates $p_k(m)$ characterises the behaviour of a \emph{measurement method}. In particular, $p_1(m)$ provides the method-level agreement prior entering the source weight, while $p_{1/2}(m)$, $p_2(m)$, $p_{1/3}(m)$, and $p_3(m)$ quantify the direction and frequency of harmonic misidentification.

%Using the unweighted plurality as the first-pass reference keeps the agreement counts independent of the weights; the prior is less protected: for a technique used by a single catalogue it reduces to that catalogue's own agreement rate. Among the catalogues that carry most of the compilation the weights span from $\approx0.99$ for the auto-correlation pipeline of \citet{McQuillan2013} to $\approx0.37$ for the lowest-agreement one, \citet{Kamai2025}. Source weight discriminates two candidate periods only when the two weights differ by at least $0.10$.

\paragraph{A-priori per-measurement reliability.}
Beyond the source weight, each measurement carries an \emph{a priori} reliability grade fixed only from its source, method, and quality flags: a measurement is \emph{low-reliability} (flagged) if it comes from deep learning, if its own catalogue reports it as weak (a near-zero autocorrelation weight, a single TESS sector, or a declared nearby contaminant), or if it carries a quality flag inherited from the source compilation; every other measurement is \emph{credible}. These grades, the source weight, and the anchor set (defined in Appendix~\ref{app:tree}) then act in the decision tree in a definite order of precedence: a credible value always beats a flagged one; between two flagged values, the source with the lower empirical halving rate, measured per catalogue, wins, with the source weight (Eq.~\ref{eq:srcweight}) breaking any tie. Where reliability does not separate the candidates, the larger source weight decides, but only when the two weights differ by at least $0.10$. Within the surviving cluster, the reported number is that of an anchor source when one is present, and the cluster median otherwise.

\subsubsection{Harmonic-alias direction and harmonic likelihoods}

\paragraph{Method-specific alias direction.}
When two candidate periods form a clean $2\times$ pair, $(P_{\rm short},P_{\rm long})$, one may be a harmonic alias of the other. The two hypotheses can be discriminated using the empirically measured method-specific rates $p_k(m)$ defined in Eq.~(\ref{eq:methodrates}), because different methods show different tendencies to underestimate or overestimate the true period. For example, the deep-learning periods of \citet{Kamai2025} predominantly return $P/2$ when they alias, whereas the gradient-of-power-spectrum method rarely produces a harmonic alias but preferentially returns a doubled period when it does \citep{Reinhold2023}.

Let $m_{\rm short}$ and $m_{\rm long}$ denote the methods associated with the two measurements. If $P_{\rm long}$ is the true period, the short measurement must represent a halving while the long measurement agrees with the reference-period scale. Conversely, if $P_{\rm short}$ is the true period, the short measurement is correctly recovered while the long measurement represents a doubling. We therefore assign the two hypotheses the relative likelihoods
\begin{align}
\mathcal{L}({\rm true}=P_{\rm long}) &=
p_{1/2}(m_{\rm short})\,p_1(m_{\rm long}), \nonumber\\
\mathcal{L}({\rm true}=P_{\rm short}) &=
p_1(m_{\rm short})\,p_2(m_{\rm long}).
\label{eq:aliaslik}
\end{align}
The favoured period is adopted when its relative confidence
\begin{equation}
{\rm conf} =
\frac{\max(\mathcal{L}_{\rm long},\mathcal{L}_{\rm short})}
     {\mathcal{L}_{\rm long}+\mathcal{L}_{\rm short}}
\end{equation}
reaches $0.70$. Below this threshold, the criterion declines to decide and the case passes to the next rule in the ranked sequence of Appendix~\ref{app:tree}. A clean $3\times$ pair is treated analogously, using $p_{1/3}(m)$ and $p_3(m)$ in place of the halving and doubling rates.
%This confidence treats the two candidate periods as the only hypotheses and the two measurements as independent. It should therefore be interpreted as a relative ranking of the hypotheses rather than as a calibrated posterior probability. As emphasised above, the rates $p_k(m)$ are internally calibrated against the first-pass reference periods of the present compilation. Consequently, the resulting confidence quantifies the relative support for the competing harmonic hypotheses within this empirical framework and should not be interpreted as a calibrated posterior probability that either candidate is the true rotation period.
This confidence treats the two candidate periods as the only hypotheses and the two measurements as independent. It should therefore be interpreted as a relative ranking of the hypotheses rather than as a calibrated posterior probability.

When the two members of a clean $2\times$ pair are obtained with the same method $m$, Eq.~(\ref{eq:aliaslik}) reduces to a comparison between that method's own halving and doubling rates, $p_{1/2}(m)$ and $p_2(m)$. The criterion can therefore still discriminate between the two hypotheses when the method exhibits a preferred alias direction; when the two rates are comparable, it declines to decide and the case passes to the source-weight criterion.

%%%%%%%%%%%%%%%%%%%%%%%%%%%%%%%%%%%%%%%%%%%%%%%%%%%%%%%%%%%%%%%%%%%%%

\paragraph{Harmonic fitting for chains of clusters.}
When three or more clusters survive, the $2\times$ logic of Eq.~(\ref{eq:aliaslik}) is generalised. 
For each candidate fundamental $C_{t}$, taken in turn as each cluster median, the relative likelihood is
\begin{equation}
\mathcal{L}(C_t) =
\prod_c
\left[
\prod_{m\in c}
p_{k_c}(m)
\right]^{1/n_c},
\qquad
k_c=\frac{\widetilde P_c}{C_t},
\label{eq:harmonic}
\end{equation}
where $\widetilde P_c$ and $n_c$ are, respectively, the median period and the number of measurements in cluster $c$. The appropriate $p_{k_c}(m)$ is selected from the method-specific rates of Eq.~(\ref{eq:methodrates}) when $k_c$ lies within $15\%$ of one of the allowed harmonic ratios $\{1/3,1/2,1,2,3\}$. Otherwise, the corresponding factor is set to $0.01$ rather than to an empirically measured rate.

The geometric mean within each cluster prevents clusters containing many, potentially correlated measurements of the same underlying light curve from receiving disproportionate statistical weight. A non-harmonic cluster therefore contributes a factor $0.01$ irrespective of its number of members and strongly disfavours the corresponding candidate fundamental. The winning fundamental is accepted only when it carries at least $0.70$ of the total likelihood and explains at least $75\%$ of the clusters as clean harmonics. Because the candidate fundamentals are restricted to the observed cluster medians, a fundamental period that is not represented by any source cannot be recovered; in such cases no candidate satisfies the coverage requirement and the case is passed on unresolved.
%the product over clusters $c$ of the geometric mean, over the $n_{c}$ members of $c$, of each member's method probability of producing the observed ratio,
%\begin{equation}
%\mathcal{L}(C_{t}) = \prod_{c} \left(\,\prod_{m\,\in\,c} P_{{\rm method}(m)}\!\left(\frac{\tilde P_{c}}{C_{t}}\right)\right)^{1/n_{c}}\,,
%\label{eq:harmonic}
%\end{equation}
%where $\tilde P_{c}$ is the median of cluster $c$. When the ratio $\tilde P_{c}/C_{t}$ falls outside the harmonic set
% $\{1/3,1/2,1,2,3\}$ (by more than $15\%$) the factor $P_{{\rm method}(m)}$ is floored at $0.01$ instead of taking a measured rate; since every member of a cluster shares the same ratio, one non-harmonic cluster drives its geometric mean to $0.01$ and strongly disfavours the candidate. The winning fundamental is accepted only when its confidence reaches the $0.70$ threshold and it explains $\ge75\%$ of the clusters as clean harmonics. The geometric mean is conservative, since members of a cluster often share the same light curve. Because candidates are drawn from the observed cluster medians, a fundamental that no source reported cannot be recovered: no candidate then reaches the coverage threshold, and the case is passed on unresolved.

%%%%%%%%%%%%%%%%%%%%%%%%%%%%%%%%%%%%%%

\subsubsection{Projected-velocity consistency}
The $P_{\max}$ ceiling only vetoes periods that are too long; a final pass over the resolved values closes the short side. 
For a transiting host whose stellar spin axis is approximately aligned with the orbital axis, $i_{\star}\simeq 90^{\circ}$  and hence $\sin i\approx1$, Eq.~(\ref{eq:pmax}) reads as the expected period $P(v\sin i)$ rather than a ceiling: a value far below it is inconsistent with the alignment and is read as a short harmonic.
When the chosen value falls below $0.55\,P(v\sin i)$ and a credible (not low-reliability) measurement, alone or within a cluster, sits at $0.7$--$1.4\,P(v\sin i)$, the value is switched to that cluster or measurement.

\subsubsection{Last-resort resolution}
Stars that no criterion resolves are retried at lower confidence: first by source trust, the a-priori grades of source and method defined above; then by alias direction at a relaxed confidence ($0.55$); then by cluster plurality; and last by a bare ranking of sources. The value is emitted as low-confidence, recording the level that resolved it; a star that even the ranking cannot resolve is released without a period (Appendix~\ref{app:outcome}).

\subsection{The rotation-period decision tree}
\label{app:tree}
The per-star decision logic runs on the measurements that survive the checks and limits of Appendix~\ref{app:guards}, branching on their number $N$; where only one survives, it is adopted directly.

\subsubsection*{The $N=2$ branch}

For a star with two surviving measurements, the rules below are evaluated in order, and the first that applies decides the value.

\begin{enumerate}
\item \textbf{Concordant pair:} if the relative difference is $\le 15\%$ (the concordance tolerance), the value is the unweighted mean of the couple.
\item \textbf{Anchor pair:} if the ratio is $\le1.30$ (discordant, but not aliases) and exactly one member comes from an anchor source (the value-setters defined below), that value is taken outright.
\item \textbf{Near-concordant pair:} a near-concordant pair (ratio $\le1.30$) that is not an anchor pair is reduced to the source-weighted mean (Eq.~(\ref{eq:srcweight})).
\item \textbf{Projected-velocity ceiling:} if exactly one member exceeds $1.3\,P_{\rm max}$ it is rotationally impossible and is discarded, keeping the other.
\item \textbf{Credible versus flagged:} if exactly one member is flagged low-reliability (Appendix~\ref{app:levers}) and the other is credible, the credible value wins.
\item \textbf{Both flagged:} if \emph{both} members are low-reliability, the less alias-prone source wins, by the empirical per-source halving-rate ranking (Appendix~\ref{app:levers}); when the two share the same rank, the decision falls back to the source weight.
\item \textbf{Harmonic pair:} if the two values stand in a clean $2\times$ or $3\times$ ratio (the measured ratio within $\sim$$15\%$ of the integer, the same window as Eq.~(\ref{eq:harmonic})), the pair is resolved by the method-specific alias direction (Appendix~\ref{app:levers}) through a combined two-hypothesis likelihood, accepted only when the confidence reaches $0.70$.
\item \textbf{Source weight:} if the two empirical source weights (Eq.~(\ref{eq:srcweight})) differ by at least $0.10$, the heavier source wins.
\item \textbf{Ambiguous pair:} if no criterion discriminates, the heavier-weighted value is passed on, to be confirmed or discarded by the last-resort criteria of Appendix~\ref{app:levers}.
\end{enumerate}

\subsubsection*{The $N\ge3$ branch}

For three or more surviving measurements, the set's fractional range is $\mathrm{spread}=(\max-\min)/\mathrm{median}$. The branch first tries two consensus scenarios in order:
\begin{enumerate}
\item \textbf{Consensus:} no estimate lies more than $25\%$ from the median; the value is the median.
\item \textbf{Outlier-rejected majority:} estimates more than $25\%$ from the median are rejected; if the survivors are more than half of the measurements and their spread is $\le15\%$, the value is their median.
\end{enumerate}

If none of these applies, the values do not form a single consensus: most often they are bimodal or aliased. They are then split into clusters wherever two consecutive periods differ by more than a factor $1.30$. Clusters violating the $P_{\rm max}$ ceiling are left out, unless they all do, in which case the shortest is kept; each of the rest is scored by its summed source weight, down-weighted by $0.30$ if all its members are low-reliability. The branch then resolves in order:

\begin{enumerate}
\item \textbf{Cluster resolved:} the top-scoring cluster beats the runner-up by $\ge1.3\times$ in summed source weight.
\item \textbf{Cluster alias:} when the top two clusters stand in a clean $2\times$ ratio in period and one of them is the cluster nearest the overall median, they are resolved by alias direction (Appendix~\ref{app:levers}) at the same $0.70$ confidence as the pair.
\item \textbf{Harmonic chain:} a chain of three or more clusters is resolved by the harmonic fit of Eq.~(\ref{eq:harmonic}), with the fundamental required to explain $\ge75\%$ of the clusters as clean harmonics, at the same $0.70$ confidence.
\item \textbf{Ambiguous cluster:} if nothing discriminates, the value at the cluster nearest the overall median is passed on; as in the ambiguous pair, the  last-resort criteria then confirms or discards it (Appendix~\ref{app:levers}).
\end{enumerate}
Where the $P_{\rm max}$ ceiling would leave no cluster at all, the shortest is used anyway and left to the vetting stage (Appendix~\ref{app:gatetiers}).

\paragraph{The cluster representative value.}
When a cluster must be reduced to one value, its low-reliability members are set aside unless it has nothing else, and the value is the median of the anchor sources among those left, or of all of them if there is no anchor:
\begin{equation}
P_{\rm rep} =
\begin{cases}
\mathrm{median}(\text{anchor members}) & \text{if any},\\[2pt]
\mathrm{median}(\text{surviving members}) & \text{otherwise.}
\end{cases}
\end{equation}

The anchor set comprises \citet{Santos2019,Santos2021}, \citet{McQuillan2013Mdwarf,McQuillan2013,McQuillan2014}, \citet{Reinhold2020}, and \citet[][hereafter SM2015]{SuarezMascareno2015}. While the others employ photometric pipelines resting on the auto-correlation function, a wavelet decomposition, or the agreement of these with a generalised Lomb--Scargle periodogram, SM2015 works on the chromospheric \ion{Ca}{ii}~H\&K and H$\alpha$ time series. A value-setter (the source that fixes the reported number) must satisfy three conditions: it measures the period independently rather than re-quoting it; it uses a method or an observable robust to harmonic aliasing (auto-correlation, wavelet, or chromospheric indices); and it agrees with the consensus at high rate (a source weight $\ge0.90$, Eq.~(\ref{eq:srcweight})).

A source can be trusted to \emph{break an alias tie} without being trusted to \emph{set the value}: the Gaussian process of \citet{Angus2018} models the quasi-periodic spot signal and, in their validation on simulated light curves, recovers the period more accurately than either the auto-correlation function or a Lomb--Scargle periodogram, so an Angus measurement can decide between aliased clusters. Its posteriors, however, are often broad or multimodal, so it does not set values: when its cluster is adopted, the reported value defers to an anchor source within the cluster or, lacking one, to the cluster median. The reliability tiers of Appendix~\ref{app:gatetiers} build on this distinction.

\subsection{Uncertainties, quality flags, and outcome}
\label{app:errors}\label{app:qflag}\label{app:manual}\label{app:outcome}

Once a single period $P_\star$ has been chosen, 
% its $1\sigma$ uncertainty is recomputed through Eq.~(\ref{eq:comb_error}) on the support set of the final value (Appendix~\ref{app:design}); 
its $1\sigma$ uncertainty is computed from Eq.~(\ref{eq:comb_error}) on the support set of $P_\star$ (Appendix~\ref{app:design});
a period resting on a single distinct value carries an additional $5\%$ error floor. Where the support set yields neither a scatter nor a quoted uncertainty, the period is published without a bar ($142$ stars) and the star receives no age.
Dedicated flags mark the periods backed only by weak measurements, those resting on a single measurement, and the values forced by the bare source ranking. Flagged values remain usable, but may carry a larger systematic uncertainty than the formal one suggests.

A small number of residual cases are resolved manually on the basis of documented literature information, with the adopted decision recorded in the published table. 
These comprise eight manually selected periods, three additional periods retained but flagged as low-confidence, and five manually excluded objects: the three Herbig\,Ae/Be stars discussed in Appendix~\ref{app:guards}, whose reported periodicities are associated with circumstellar-disc variability, the red giant Kepler-56, and Kepler-1617, for which none of the candidate periods could be confirmed.

Overall, of the $2874$ hosts for which a rotation period is reported in the input literature, $158$ retain no direct rotation-period measurement after the initial cleaning described in Appendix~\ref{app:guards}. Of the remaining stars, $10$ are excluded during the subsequent selection and vetting procedure because of the break-up limit, documented manual exclusions, or mutually discordant measurements that cannot be resolved by any of the adopted criteria. The procedure therefore yields an adopted rotation period for $2706$ stars ($\approx94\%$ of the initial sample), of which $52$ are flagged as low-confidence.

\subsection{Stellar-mass selection}
\label{app:mass}

The mass prior $\mathcal{M}_0\pm\sigma_{\mathcal{M}}$ of Sect.\,\ref{sec:mass} follows the same scheme as the period but, with no aliasing to resolve, reduces to a consensus selection, as detailed below.
Each star is first assigned to an operational stellar regime using its selected $\log g$ and radius (Appendix~\ref{app:loggradius}), solely for the purpose of applying the appropriate parameter-selection criteria. Stars are assigned to the \emph{giant regime} when $\log g<3.5$ or $R>3.5\,R_\odot$, and to the \emph{M-dwarf regime} when $R<0.6\,R_\odot$ or $\log g>4.6$; all remaining stars are assigned to the \emph{FGK-dwarf regime}. These definitions are not intended as spectral classifications, but provide a practical separation between regions of parameter space in which the reliability and applicability of the available mass determinations differ. 
Two pre-filters are then applied: Kepler Object of Interest placeholder masses (quoted with zero uncertainty) are discarded whenever a genuine estimate is available, while for microlensing hosts the FLAME mass \citep{Creevey2023} is excluded because it may refer to the source star rather than to the lens.
With the surviving estimates the algorithm applies an ordered sequence of rules, stopping at the first that fixes the value, with the error following the conservative rule of Eq.~(\ref{eq:comb_error}). The sequence depends on the number $N$ of surviving estimates:
\begin{itemize}
  \item \emph{$N=1$}: the lone estimate is adopted; its trustworthiness is left to the tier assignment.
  \item \emph{$N=2$}: if concordant within $15\%$ ($20\%$ for M~dwarfs), the value is the mean of the two. Failing that, the estimate from the more precise method is adopted, the precision of a method being its median fractional scatter about the per-star median, measured once on the stars with at least three measurements. Where the two scatters agree to within $0.02$ the smaller relative uncertainty decides instead, and where the two relative uncertainties agree to within $0.01$ the chosen value is the mean, with the error increased to half the gap.
  \item \emph{$N\ge 3$}: when no estimate lies more than $35\%$ from the median, the value is the median; otherwise, the estimates beyond that threshold are discarded and, if a majority survives, the value is their median; any residual case is set as the median, with the uncertainty taken as the larger of Eq.~(\ref{eq:comb_error}) and half the spread. The threshold is looser than rotation's $25\%$: with no aliases to keep out, a wider window absorbs the genuine model-to-model spread.

\end{itemize}

On top of this generic sequence, two physically motivated rules take priority where they apply.

\smallskip\noindent\textit{(1) Giants.}
For evolved stars we keep only asteroseismic, spectroscopic, isochrone, and dynamical masses, in that order, or, as a last resort, the mass recomputed from the measured $\log g$ (or mean density) and the radius. All estimates that assume the main sequence are discarded. A giant with none of the reliable methods is excluded.

\smallskip\noindent\textit{(2) Microlensing.}
A microlensing host is kept when it has a single value or solutions concordant within the $15\%$ tolerance, and excluded when they disagree by more, since such a spread may reflect degenerate solutions.

\subsection{Colour and de-reddening}
\label{app:colour}

The colour $C_\star$ of Sect.\,\ref{sec:colour} is obtained by \emph{de-reddening} the default
{\it Gaia} DR3 $G_{\rm BP}-G_{\rm RP}$ measurement of each star, $C_{\star,{\rm obs}}$
\citep[][with its formal error $\sigma_{C,{\rm obs}}$]{GaiaDR3}:
\begin{equation}
C_\star = C_{\star,{\rm obs}} - E(G_{\rm BP}-G_{\rm RP}),
\end{equation}
with $E(G_{\rm BP}-G_{\rm RP})$ read from the {\it Gaia}-based three-dimensional extinction map of \citet{Vergely2022}, integrated along the line of sight to each star's adopted distance, the most precise of its estimates among {\it Gaia}~DR3 GSP-Phot \citep{GaiaDR3}, \citet{BailerJones2021}, TICv8 \citep{Stassun2019}, and \citet{Berger2020,Berger2026}. For sight lines that exit the mapped volume the extinction is integrated only to the boundary, a lower bound; these stars are a small fraction of the final sample ($56$ of the $2204$, $\sim$$2.5\%$) and carry the lower-reliability extinction tier ($2$, against $1$ in-volume). Stars with no map value at all, or no {\it Gaia} colour, are \emph{excluded} from the sample. The map is tabulated as an extinction density in $A_V$ and converted with $E(G_{\rm BP}-G_{\rm RP})=0.413\,A_V$\footnote{$0.413\,(\pm0.008) = A_{G_{\rm BP}}/A_V - A_{G_{\rm RP}}/A_V$, with $A_{G_{\rm BP}}/A_V = 1.002\,(\pm0.007)$ and $A_{G_{\rm RP}}/A_V = 0.589\,(\pm0.004)$ fixed by the \citet[][Table~3]{WangChen2019} extinction-law shape ($R_V=3.16\pm0.15$).}.

The reddening uncertainty is added in quadrature, $\sigma_C=\sqrt{\sigma_{C,{\rm obs}}^2+\sigma_E^2}$. The reddening error folds three terms, $\sigma_E=\sqrt{(0.413\,\sigma_{A_V}^{\rm formal})^2+\left(\tfrac{\mathrm{d}E}{\mathrm{d}d}\,\sigma_d\right)^2+(A_V\,\sigma_{0.413})^2}$: the formal per-sightline posterior uncertainty of the extinction map $\sigma_{A_V}^{\rm formal}$, already incorporating the uncertainties of the input extinction catalogue from which the map is built; the distance term, the local line-of-sight reddening gradient $\mathrm{d}E/\mathrm{d}d$ times the distance uncertainty $\sigma_d$; and the propagated uncertainty of the conversion coefficient, $\sigma_{0.413}=0.008$ \citep{WangChen2019}. The first two are random per star, the third a correlated systematic carried conservatively. The correction is small (median $E\simeq0.05$\,mag), so $\sigma_C$ is set mainly by $\sigma_{C,{\rm obs}}$, not $\sigma_E$.

\subsection{Auxiliary stellar parameters}
\label{app:aux}

Four further quantities are selected, but do not enter the gyrochronological likelihood: the effective temperature $T_{\rm eff}$, metallicity $\mathrm{[Fe/H]}$, surface gravity $\log g$, and radius $R$. For all four the dominant systematic is the measurement method, and the estimates are combined by the same consensus rules as the mass.

\subsubsection{Effective temperature and metallicity}
\label{app:tefffehcolour}

The inputs are ranked into four method tiers of decreasing reliability: interferometry, the temperature following from the angular diameter and the bolometric flux; spectroscopy, essentially immune to reddening as it rests on line strengths rather than on the spectral energy distribution; isochrone fitting, combining photometry, parallax, and stellar-evolution models; and photometric estimates, which are entangled with the reddening solution.
Each star is classified into the regimes of Appendix~\ref{app:mass} and descends the hierarchy until a value exists. For M~dwarfs, near-infrared spectroscopy comes ahead of the optical, which their molecular bands veil. The photometric fallback is flagged unreliable for M~dwarfs (molecular opacity) and giants (granulation), and accepted at the lowest tier for FGK dwarfs.
Within the selected tier, the values are combined following the same consensus procedure adopted for the mass, while their uncertainties are assigned according to Eq.~(\ref{eq:comb_error}). The consensus tolerance is $3\%$ for $T_{\rm eff}$ ($5\%$ for stars in the M-dwarf regime) and $0.10$\,dex for $\mathrm{[Fe/H]}$ ($0.15$\,dex for stars in the M-dwarf regime). Measurements more than $10\%$ from the median in $T_{\rm eff}$, or $0.50$\,dex in $\mathrm{[Fe/H]}$, are rejected when the remaining values constitute a majority; if no such majority can be identified, the median of all values is retained, with its uncertainty taken as the larger of Eq.~(\ref{eq:comb_error}) and half the full spread.
The selected $T_{\rm eff}$ is spectroscopic for $\sim81\%$ of the sample, and the selected $\mathrm{[Fe/H]}$ for $\sim76\%$, with isochrone-based determinations supplying most of the remaining values.

\subsubsection{Surface gravity and radius}
\label{app:loggradius}

$\log g$ and $R$ follow separate hierarchies. For the gravity: dynamical modelling (photodynamical fits, astrometric orbits, eclipsing binaries) first, then asteroseismology, isochrone fitting, spectroscopy, and lastly photometric estimates, with the {\it Gaia} GSP-Spec gravities excluded: they are systematically lower, the more so for giants \citep{RecioBlanco2023}, and depart from the other spectroscopic determinations of the same star by more than their mutual spread.
For the radius: interferometry first, then dynamical modelling, asteroseismology, isochrone fitting, the Stefan--Boltzmann radius (from luminosity and temperature, as in the {\it Gaia} FLAME radius), and lastly photometric estimates; for giants the Stefan--Boltzmann tier outranks the isochrone one.
The within-tier combination follows the temperature case, with a consensus tolerance of $5\%$ for the radius and $0.10$\,dex for $\log g$ ($10\%$ and $0.15$\,dex for M~dwarfs). A single exception applies: when two values in the winning tier differ by a factor of $\sim$2, typically a dwarf against a giant solution of the same star, the smaller error bar does not decide; the value corroborated by the other tiers does.

\subsubsection{Cross-parameter coherence}
\label{app:coerenza}

As a final cross-parameter consistency check, the separately selected mass, radius, and surface gravity are compared through
\begin{equation}
\log g_{\mathcal{M},R}=4.438+\log(\mathcal{M}/\mathcal{M}_\odot)-2\log(R/R_\odot)\,.
\end{equation}
If the selected $\log g$ differs from $\log g_{\mathcal{M},R}$ by more than $0.3$\,dex, all the gravity measurements available for that star are reconsidered and the highest-ranked alternative that reduces the discrepancy to less than $0.1$\,dex is adopted. 
The $0.3$\,dex threshold is used as a deliberately conservative empirical criterion to identify substantial inconsistencies among separately selected parameters, whereas the stricter $0.1$\,dex tolerance is required for a replacement to be considered genuinely consistent with the adopted mass and radius. 
This re-selection is not applied when the adopted $\log g$ is based on asteroseismology methods (2 cases), since in that case forcing agreement with $\mathcal{M}$ and $R$ could replace an independent constraint with one effectively derived from the other selected parameters. If no acceptable alternative is available, the original $\log g$ is retained and the star is flagged as cross-parameter inconsistent.

\subsection{Quality flags and reliability tiers}
\label{app:gatetiers}

After the selection a final vetting stage reads the parameters together and, without altering any value, attaches the quality flags and reliability tiers.
The period carries two kinds of flags: \emph{data-quality} (a weak detection, a lone low-reliability source, an uncertain resolution) and \emph{physics} (a period rotationally impossible for the implied $v\sin i$, or anomalously long for the star's mass\footnote{Above $70\%$ of the turnoff period of the track at that mass, only for stars bluer than $C_\star=0.85$. Near the turnoff the expanding radius drives a steep late rise in period, collapsing a broad band of long periods onto a near-degenerate age; this singles out late-F and early-G stars near the Kraft break.}).

The adopted rotation periods are assigned a reliability tier that combines the outcome of the selection procedure with the quality and independence of the available measurements. 
Tier~1 is reserved for unflagged periods resolved by mutually consistent measurements, by an anchor source, or by the $v\sin i$ constraint, demoted to Tier~2 where the supporting agreement rests on a single facility. Tier~2 comprises the remaining unflagged periods. Tier~3 identifies periods affected by data-quality concerns: a weak detection, an uncertain resolution, or a lone measurement from a low-reliability source. Tier~4 is reserved for periods affected by physical-consistency concerns, which take precedence where both kinds of flag are present. 
In addition, for hot stars with $C_\star<0.85$, a disagreement exceeding $20\%$ between the adopted period and the period predicted by the rotational model is treated as a post-inference physical-consistency flag and demotes the period to Tier~4. Among the 2204 stars for which a gyrochronological constraint is obtained, 246 rotation periods are assigned to Tier~1, 1146 to Tier~2, 390 to Tier~3, and 422 to Tier~4.

Reliability tiers are also assigned independently to the other adopted stellar parameters. For mass, $T_{\rm eff}$, $\mathrm{[Fe/H]}$, $\log g$, and radius, the tier reflects the quality of the underlying measurement method, the degree of agreement among the available determinations, and any parameter-specific flags introduced by the selection procedure. The de-reddened Gaia colour is similarly assigned a reliability tier according to the availability and quality of the observed photometry and reddening correction. These parameter-level tiers therefore describe the reliability of the adopted catalogue quantities and are retained independently of the subsequent gyrochronological inference.

The reliability tier of the inferred age, $q_{\rm Age}$, is defined as the least favourable of the tiers assigned to the three quantities entering the gyrochronological inference: the rotation period, stellar mass, and de-reddened colour $C_\star$. It is therefore propagated directly from the quality of the observational constraints rather than from the width of the age posterior. Two further consistency checks on the adopted parameters demote $q_{\rm Age}$ to Tier~4: a de-reddened colour bluer than the blue edge of the model grid at the adopted mass, and an inconsistency among the selected $\log g$, mass, and radius according to the cross-parameter check described above. Finally, inferred ages exceeding $13.8$\,Gyr are assigned to Tier~4 as a physical-consistency veto; this affects 13 stars. Thus, $q_{\rm Age}$ should be interpreted primarily as a reliability indicator for the information entering the gyrochronological inference, supplemented by explicit physical-consistency flags, rather than as a measure of the formal precision of the posterior age.

%%%%%%%%%%%%%%%%%%%%%%%%%%%%%%%%%%%%%%%%%%%%%%%%%%%%%%%%%%%%%%

\end{appendix}

\end{document}